\documentclass[aps,nofootinbib,showpacs,showkeys,preprint]{revtex4}

\usepackage{epsf,epsfig,subfigure,graphicx,amsmath,amssymb}

\def\sr2{\sqrt{2}}

\def\to{\rightarrow}

\def\bi{\begin{itemize}}
\def\ei{\end{itemize}}

\def\c1p{C1^\prime}
\def\ta{\tilde a}
\def\tG{\widetilde G}

\def\tu{\tilde u}

\def\ta{\tilde a}

\def\tst{\tilde t}

\def\tg{\tilde g}

\def\tz{\widetilde Z}

\def\be{\begin{equation}}
\def\ee{\end{equation}}
\def\bea{\begin{eqnarray}}
\def\eea{\end{eqnarray}}

\newcommand\prD[3]{{Phys.\ Rev.\ }{\bf D#1} (#2) #3}
\newcommand\prL[3]{{Phys.\ Rev.\  Lett.\ }{\bf #1} (#2) #3}

\def\Isajet{{\sc Isajet}}

\begin{document}

\title{Mixed axion/gravitino dark matter\\
from SUSY models with heavy axinos
}
\author{Kyu Jung Bae$^{a,b}$\footnote{bae@nhn.ou.edu}, 
Howard Baer$^{a,b}$\footnote{baer@nhn.ou.edu},
Eung Jin Chun$^{c}$\footnote{ejchun@kias.re.kr} and Chang Sub Shin$^{d}$
\footnote{changsub@physics.rutgers.edu}}
\affiliation{
$^a$Dept.\ of Physics and Astronomy, University of Oklahoma, Norman, OK 73019, USA\\
$^b$William I. Fine Theoretical Physics Institute, University of Minnesota, Minneapolis, MN 55455, USA\\
$^c$Korea Institute for Advanced Study, Seoul 130-722, Korea\\
$^d$New High Energy Theory Center, Department of Physics and Astronomy,\\
Rutgers Unversity, Piscataway, NJ 08854, USA\\
}

\begin{abstract}
We examine dark matter production rates in supersymmetric axion models typified by the mass 
hierarchy $m_{3/2}\ll m({\rm neutralino})\ll m({\rm axino})$. In such models, one expects the dark 
matter to be composed of an axion/gravitino admixture. After presenting motivation for how such
a mass hierarchy might arise, we examine dark matter production in the SUSY Kim-Shifman-Vainshtein-Zakharov (KSVZ) model, the
SUSY Dine-Fischler-Srednicki-Zhitnitsky (DFSZ) model and a hybrid model containing contributions from both KSVZ and DFSZ. 
Gravitinos can be produced thermally and also non-thermally from axino, saxion or neutralino 
decay. We obtain upper bounds on $T_R$ due to overproduction of gravitinos including both the thermal and non-thermal processes.
For $T_R$ near the upper bound, then dark matter tends to be gravitino dominated, but for $T_R$
well below the upper bounds, then axion domination is more typical although in many cases
we find a comparable mixture of both axions and gravitinos.
In this class of models, we ultimately expect detection of relic axions but no WIMP signal, 
although SUSY should ultimately be discovered at colliders.
\end{abstract}



\preprint{FTP-MINN-14/31}
\preprint{RUNHETC-2014-17}

\maketitle

\section{Introduction}

The axion solution to the strong CP problem provides a natural candidate for dark matter, 
the cold axion produced
coherently from an initial misalignment during the QCD phase transition \cite{Kim08}.  
In its supersymmetric (SUSY) version, the axion, $a$, is accompanied by the fermionic and scalar partners called the axino, $\tilde a$,  and saxion $s$, respectively. They also have significant implications in cosmology \cite{Raja91,Baer:2014eja}, which are 
characteristically different depending on the axion models. 
In the Kim-Shifman-Vainshtein-Zakharov (KSVZ) model \cite{ksvz}, the axion solution is realized by the presence of extra heavy vector-like quarks and thus 
the axion supermulitiplet interacts with the minimal supersymmetric standard model (MSSM) fields through 
the (non-renormalizable) QCD anomaly term. 
On the other hand, in the Dine-Fischler-Srednicki-Zhitnitsky (DFSZ)  model \cite{dfsz}, the  $\mu$-problem of the MSSM is connected to the axion solution \cite{KimNilles} and the (renormalizable Yukawa-type) $\mu$-term interaction plays a major role in the axino/saxion cosmology.

Since the axion is the Goldstone boson of a spontaneously broken U(1)$_{PQ}$ symmetry \cite{pqww}, 
its mass is protected to be zero up to the QCD anomaly.  The axino and saxion remain also massless in the SUSY limit.  
In reality, however, SUSY breaking induces their masses which are generically expected to be of order 
the SUSY breaking scale, but can be quite model-dependent \cite{Goto92,Chun92,Chun95,Chun99,ks}.
Being superpartners of a Goldstone boson, the axino and saxion interact 
with the MSSM particles through couplings suppressed by the axion scale $f_a \sim 10^9 - 10^{12}$ GeV.
Although very weakly coupled, sizable cosmic abundances of the axino and saxion 
can be generated either through the QCD anomaly interaction in the KSVZ model \cite{Covi01,Brand04,Strumia10}, 
or  through the $\mu$-term interaction  in the DFSZ model \cite{Chun11,bci1,bci2}.  
Thus, the axino has to be very light if it is the lightest supersymmetric particle (LSP) and thus a dark matter candidate 
\cite{cckr11}.  If the axino (or saxion) is heavy and unstable, its decay leads to a large non-thermal abundance of the LSP
such as a neutralino or the gravitino, which can change the standard dark matter cosmology significantly.
Note also that coherent oscillation (CO) is another important source of the saxion cosmic abundance.

If the gravitino is the LSP,  its abundance comes from the usual thermal generation depending on the reheat temperature
and also from non-thermal generation due to next-to-lightest SUSY particle (NLSP) decays. 
This contribution is important if the axino is the NLSP 
due to its sizable initial abundance \cite{Cheung11}. If a usual neutralino is the NLSP,  the axino (and saxion) typically decays
first to the NLSP and then its re-adjusted abundance will be relevant to the 
gravitino production while the direct decay of the 
axino (saxion) to the gravitino is suppressed by $ {\cal O}(m_{\ta, s}^2 f_a^2/m_{3/2}^2M_P^2)$ 
which is a tiny number for $m_{\ta, s} \sim m_{3/2}$.
In this paper, we investigate the  possibility of realizing the situation that the axino/saxion mass 
is  hierarchically larger than the gravitino mass and thus the axino/saxion decay to the gravitino cannot be negligible.

In Section 2, we first consider the effective theory of the axion supermultiplet to see how rather 
unusual cases of $m_{\ta, s} \gg m_{3/2}$ can be realized and then provide specific examples in gravity and gauge mediation models.
In Section 3, some phenomenological implications of SUSY KSVZ and DFSZ axion models will be discussed.
If $m_{3/2} \ll 100$ GeV, the SUSY breaking masses in the MSSM sector can be generated by the 
usual gauge mediation or the ``axionic gauge mediation'' which can be realized in the KSVZ scheme. 
In Section 4, we investigate the cosmological consequences of heavy axinos/saxions by 
taking specific examples of the Higgsino-like (SUA) and bino-like (SOA) NLSP. 
For these benchmark points, we compute the gravitino abundance coming
from thermal generation\cite{Khlopov:1984pf}, the NLSP and axino/saxion decays, and put an upper bound on the 
reheat temperature in the  KSVZ\cite{Graf:2013xpe}, DFSZ and hybrid (KSVZ+DFSZ) axion models.  
We also present a brief discussion on the big-bang nucleosynthesis (BBN) bound on the long-lived NLSP.  
Finally, we conclude in Section 5.

\section{Axino and saxion masses in effective theory}
\label{sec:mass}

The main focus of this paper is to  investigate the consequences of a rather exotic mass spectrum:
\bea
m_{3/2} \ll m_{\widetilde{Z}_1} \ll m_{\tilde a},
\eea
which can be realized  in both gravity mediation and gauge mediation scenarios.
To see how this happens, let us revisit the effective theory\cite{Higaki:2011bz} 
of the axion supermultiplet $A$,
\begin{equation}
A=\frac{1}{\sqrt{2}}\left(s+ia\right)+\sqrt{2}\theta\tilde{a}+\theta^2F_A,
\end{equation}
which is a Goldstone superfield arising after spontaneous breaking of  U(1)$_{PQ}$ symmetry in SUSY theory.
The low energy effective theory below the Peccei-Quinn (PQ) symmetry breaking scale $v_{PQ}$
should be invariant under the non-linear transformation of $A$:
\bea
{\rm U(1)}_{PQ}:\quad A\to A+ i\alpha\, v_{PQ},
\eea
where $\alpha$ is a real parameter, and other fields are all neutral under U(1)$_{PQ}$.
In order to be invariant under U(1)$_{PQ}$, the effective superpotential $W_{\rm eff}$
should be independent of $A$, and the effective K\"ahler potential $K_{\rm eff}$
should be the function of $A+A^\dagger$.
Expanding $K_{\rm eff}$ in terms of $(A+A^\dagger)/v_{PQ}$, one has
  \bea\label{eq:axion1}
  K_{\rm eff}
  = v_{PQ}^2 \left({\cal Z}_0 +  {\cal Z}_1 \frac{(A+A^\dagger)}{v_{PQ}} +
\frac{{\cal Z}_2}{2!} \frac{(A+A^\dagger)^2}{ v_{PQ}^2} +
\frac{ {\cal Z}_3}{3!} \frac{(A+A^\dagger)^3}{v_{PQ}^3} +\cdots \right)
  \eea
where ${\cal Z}_{i}$ are spurion superfields.
Assuming that there is no significant mixing between the axino and other fermions,
${\cal Z}_{i}$ can be written as
\bea
{\cal Z}_{i} = Z_{i}
+(\theta^2 Z_{i}^F + {\rm h.c.})
+ \theta^2\bar\theta^2 Z_{i}^D.
\eea
Calculating $K_{\rm eff}|_{\theta^2\bar\theta^2}$,
and solving the equations of motion for $F_A$,  
we obtain  
\bea\label{eq:FA_term}
\frac{F_A}{v_{PQ}} = -\frac{Z_1^F}{Z_2} -\sqrt{2}\left(\frac{Z_2^F}{Z_2}  - \frac{Z_3  Z_1^F}{Z_2^2} \right)\frac{s}{v_{PQ}}. 
\eea
Here we keep terms up to ${\cal O}(1/v_{PQ})$.
Considering the scalar potential for $s$ induced by ${\cal Z}_i$ and
the constraint $\langle s\rangle =0$, one finds
\bea
\left|Z_1^F\right|^2 =  \frac{(Z_2^{F*} Z_1^F + Z_2^F Z_1^{F*} -Z_1^D Z_2) Z_2}{Z_3 }.
\eea
Barring an additional symmetry or a special arrangement,  
it is generally expected that
\bea Z_i= {\cal O}(1), \quad
Z_i^D = {\cal O}\left((Z_j^F)^2\right).
\eea
Then one can find that the axino mass is given as
\bea\label{eq:axino_mass}
m_{\tilde a} = \frac{Z_2^{F*}}{Z_2} - \frac{Z_3 Z_1^{F*}}{Z_2^2}
= \frac{Z_2^{F*}}{Z_2} +  \frac{Z_3}{Z_2}\frac{ F_A^* }{v_{PQ}}
={\cal O}\left(Z_2^F\right), \;\mbox{or}\; \; {\cal O}\left(\frac{ F_A}{v_{PQ}}\right) .
\eea
Similarly, the saxion mass-squared is
\bea
m_s^2 =2\left( \frac{2 |Z_2^F|^2}{Z_2^2}+ \frac{Z_1^F Z_3^{F*} + Z_1^{F*} Z_3^F}{Z_2^2}
 -\frac{Z_2^D}{Z_2} - \frac{2 Z_1^D Z_3}{Z_2^2}   \right)
 \sim {\cal O}(m_{\tilde a}^2).
\eea

As an example of an UV model with an additional (approximate) symmetry
$A\leftrightarrow -A$ requiring  $F_A=0$, let us introduce two PQ charged chiral superfields ($X,\, Y$)
transforming like
\bea
{\rm U(1)}_{PQ}:\quad X\to X e^{i\alpha},\quad Y\to Y e^{-i\alpha}.
\eea  They can be decomposed as
\bea
X= \frac{1}{\sqrt{2}} U e^{A/v_{PQ}},\quad Y = \frac{1}{\sqrt{2}}U e^{- A/v_{PQ}}
\eea
where $U$ is  a PQ neutral spurion superfield whose vacuum value is
determined by equations of motion. The transformation $A\leftrightarrow -A$
corresponds to $X\leftrightarrow Y$, and  $\langle X\rangle =
\langle Y\rangle =  v_{PQ}/\sqrt{2}$.
After stabilization of the $U$ field, the low energy effective K\"ahler potential is
\bea\label{eq:axion2}
K_{\rm eff} = {\cal Z}_{\rm eff}\left( X^\dagger X + Y^\dagger Y\right) =
 v_{PQ}^2
{\cal Z}_{\rm eff}\left|\frac{U}{v_{PQ}}\right|^2  \cosh
\frac{(A+A^\dagger)}{v_{PQ}},
\eea
where ${\cal Z}_{\rm eff}$ can be taken as $1+\theta^2\bar\theta^2 m_*^2$,
since $\theta^2$ ($\bar\theta^2$) terms could be removed by field redefinition of $X$, $Y$.
By matching (\ref{eq:axion2}) with (\ref{eq:axion1}), we get
\bea
{\cal Z}_1={\cal Z}_3=0\quad
{\cal Z}_0={\cal Z}_2 = 1 + \left(\theta^2 \frac{  F_U}{v_{PQ}} + {\rm h.c.}\right)
+ \theta^2 \bar\theta^2 \left(m_*^2 + \left|\frac{F_U}{v_{PQ}}\right|^2\right)
\eea
and the axino mass is
\bea
m_{\tilde a} =\frac{ F_U}{v_{PQ}}.
\eea
This corresponds to
\bea
m_{\tilde a} =  \frac{F_X}{X_0}= \frac{F_Y}{Y_0},
\eea
where $X_0\equiv\langle X\rangle$ and $Y_0\equiv\langle Y\rangle$ in the linearly realized PQ symmetry.

From the above discussion, one can get a formal upper bound for the axino mass as a function of
the gravitino mass and the PQ symmetry breaking scale.
For a given gravitino mass,  $F$-terms are bounded as $|F_A|,\ |F_U| < \sqrt{3} m_{3/2} M_{P}$ which leads to
\bea\label{eq:axn_con}
m_{\tilde a} <  m_{3/2}\Big(\frac{M_{P}}{v_{PQ}}\Big).
\eea
On the other hand, the saxion mass-squared in the above example  is
\bea
m_s^2 =2\left\{\left(\frac{ F_U}{v_{PQ}}\right)^2- m_*^2\right\}
=2 \left( m_{\tilde a}^2 - m_*^2\right) \gtrsim {\cal O}(m_{\tilde a}^2).
\eea
The specific relation between $m_{\tilde a}$ and $m_s$ is model-dependent.
Let us remark that the relation  (\ref{eq:axn_con}) allows a hierarchically large ratio $m_{\tilde a}/m_{3/2}$ up to $M_P /v_{PQ}$  which has not been studied seriously in the literature  as one generically finds $Z_2^F\sim m_{3/2}$ and  $|F_{A,U}|\sim m_{3/2} v_{PQ}$,
and thus $m_{\tilde a} \sim m_{3/2}$.

\subsection{Gravity Mediation}


 In gravity mediation models, the axino mass is of order the gravitino mass or smaller 
if the theory does not have an additional zero mode other than the axion mode in the supersymmetric vacuum.
On the other hand, if the theory does have an additional zero mode, 
the axino mass can be hierarchically larger than gravitino mass~\cite{Chun95}.
We show here a specific example realizing $m_{\tilde a}\gg m_{3/2}$ in gravity mediation models.

Let us consider the following superpotential:
\begin{equation}
W=\left(\lambda_x X Y -\lambda_z Z^2\right) S  +\lambda_f \left(Z-f_0\right)^3,
\label{eq:model}
\end{equation}
with U(1)$_{PQ}$ charges
$X (1),\ Y (-1),\ Z(0),\ S(0)$.
 In the supersymmetric limit, the vacuum expectation values of 
$X,\, Y,\, Z,\, S$ ($X_0,\, Y_0,\, Z_0,\, S_0$) are
\bea
S_0= 0,\quad Z_0= f_0,\quad X_0Y_0 = (\lambda_z/\lambda_x) f_0^2.
\eea
We find that in addition to the axion supermultiplet corresponding to the flat direction $X_0Y_0=$ constant, 
there is another massless spectrum whose mass is proportional to $(Z-f_0)$. 
This is the accidental massless mode, which does not correspond to any flat direction, 
and thus is removed if terms like $SZ$ and extra $Z^2$ are added.\footnote{
It is noted that for sizable $\lambda_z$, then  mixing terms 
between $S$ and $Z$ can be induced in the K\"aher potential. 
Such terms are quite suppressed for $\lambda_z\ll 1$, 
and our tree-level discussions are still valid at the loop level.}
The vacuum values are modified 
when the SUSY breaking terms are turned on. 
The modification of the vacuum values are ${\cal O}(m_{3/2})$ along the massive directions 
while they can be much larger than $m_{3/2}$ along the additional massless direction.
More specifically, the superpotential term $(Z-f_0)^3$ makes a large shift of $\delta Z_0\sim {\cal O}(m_{3/2}^{1/3}f_0^{2/3})$ and consequently $m_{\tilde{a}}\sim\delta S_0\sim{\cal O}(m_{3/2}^{2/3}f_0^{1/3})$.
The relation between the vacuum structure and  the mass spectrum 
before and after adding soft SUSY breaking terms is discussed more generally in Ref.~\cite{Chun95}.

To simplify the analysis, let us assume that
$\lambda_f\sim\lambda_x\sim1\gg\lambda_z$. Including generic gravity mediated soft terms in the scalar potential, we obtain the mass spectrum of the PQ sector as follows:
\begin{eqnarray}
m_a&=&0,\\
m_{s}&\simeq& \sqrt{2}\, m_{\tilde{a}}\simeq \sqrt{2} \lambda_x S_0\sim \left(\lambda_f m_{3/2}^2f_0\right)^{1/3},\\
m_{s2}&\simeq & m_{s3}\simeq m_{p2} \simeq m_{p3}\simeq m_{\tilde{p}}\simeq m_{\tilde{q}}\simeq \lambda_x X_0 \sim f_a,\\
m_{s4} &\simeq& \sqrt{3}\, m_{p4} \simeq  \sqrt{6}\, m_{\tilde{z}}\sim \left(\lambda_f^2m_{3/2}f_0^2\right)^{1/3},
\end{eqnarray}
where $\{s,\, a,\, \tilde a\}$ are the axion supermultiplet,
$\{s2,\cdots, s4\}$ are scalars, $\{p2,\cdots, p4\}$ are pseudoscalars,
and $\{\tilde p,\tilde q, \tilde z\}$ are fermion superpartners.
One can see that  $m_s,\, m_{\tilde{a}}\gg m_{3/2}$ is obtained for $f_0\gg m_{3/2}$.
We should note that in this setup every superfield has a non-zero $F$-term of order
\begin{eqnarray}\label{Fterm}
F_X/X_0\simeq F_{Y}/Y_0\simeq m_{\tilde{a}},
\qquad F_Z/Z_0\simeq -m_{\tilde{a}}\qquad F_S/S_0\simeq -m_{\tilde{a}}.
\end{eqnarray}

A peculiar feature of the model (\ref{eq:model}), which is relevant to cosmology,
is that the saxion decay to a pair of axions/axinos  is very suppressed by the small coupling $\xi$
which will be discussed later in Eqs.~(\ref{eq:Lag_ax_sup},\ref{eq:saxion_dec}) \cite{Chun95}:
\bea
\xi&=& \sum_i q_i^3 v_i^2/v^2_{  PQ} \nonumber\\ &=&
\frac{X_0^2 -Y_0^2}{X_0^2  + Y_0^2}
\nonumber\\
&=& \frac{m_{Y}^2 - m_X^2}{2m_{\tilde a}^2 + m_X^2 + m_{Y}^2}
\sim \left(\frac{m_{3/2}}{v_{PQ}}\right)^{2/3} \ll 1.
\eea
where $m_X^2$ ($m_{Y}^2$) is the soft scalar mass squared for $X$ ($Y$)
of order $m_{3/2}^2$.
In addition, the saxion decay to an axino pair is also kinematically forbidden due to 
$m_s\simeq \sqrt{2} m_{\tilde a} < 2 m_{\tilde a}$.

\subsection{Gauge Mediation}

In the usual gauge mediation model, one has $m_{3/2}\ll m_{\rm soft}$. 
One can also expect to have $m_{\tilde a} \ll m_{\rm soft}$
as the PQ symmetry breaking sector consists of gauge singlet fields \cite{Chun99}.
To get the opposite spectrum of  $m_{\rm soft}\ll  m_{\tilde a}$, one needs to allow  a direct coupling
between the axion superfield and the SUSY breaking/messenger field.\footnote{Here we consider that dominant SUSY breaking fields
are not charged under the U(1)$_{PQ}$, so that
the axion sector stabilization is independent of the SUSY breaking sector construction.}
For a given SUSY breaking spurion superfield $Z = Z_{0} + \theta^2 F_{Z}$,
we introduce $N_M$ copies of PQ charged SM singlet chiral superfields {$M+M^c$}
as  messengers between the SUSY breaking and the axion sector.
The U(1)$_{PQ}$ charges are assigned as
\bea
X(1),\, Y(-1),\, Z(0),\, M(-1/2),\,  M^c(1/2), \, S(0).
\eea
The PQ invariant superpotential is
\bea
W&=& Z \Phi\Phi^c+  \lambda Z M M^c  \nonumber\\
&&+\, {1\over2} \kappa_x X M M + {1\over2} \kappa_{y} Y M^c M^c  \nonumber\\
&&+\, (\lambda_x X Y - f_0^2) S.
\eea
where $\Phi+\Phi^c$ are SM charged messenger superfields and the
first term $Z\Phi\Phi^c$ is the source of gauge mediation for the MSSM sector.
The coupling $XYZ$ can be prevented by assigning additional $U(1)_R$ charges.
For $f_a\sim X_0\lesssim Z_0$,
$M+M^c$ are integrated out at the scale $Z_0$.
At one-loop level, this effect can be captured by the Coleman-Weinberg K\"ahler potential as
\bea
\Delta K_{\rm eff}= -\frac{1}{32\pi^2} {\rm Tr}\left({\cal M}^\dagger {\cal M}\ln\frac{{\cal M}^\dagger {\cal M}}{\Lambda^2}\right),
\eea
where ${\cal M}$ is the mass matrix for $M$ and $M^c$ that depends on $X$, $Y$, $Z$.
Then we have
\bea
\Delta K_{\rm eff} &=&  -\left(\frac{N_M\,\kappa_x^2}{32\pi^2}\ln\frac{\lambda^2|Z|^2}{\Lambda^2}\right)|X|^2 -\left(\frac{N_M\,\kappa_y^2}{32\pi^2}\ln\frac{\lambda^2|Z|^2}{\Lambda^2}\right)|Y|^2 \nonumber\\
&&-\, \frac{N_M\,(\kappa_x^2|X|^4 + \kappa_y^2|Y|^4)}{64\pi^2 \lambda^2|Z|^2} +\cdots
\eea
Taking $\kappa_x=\kappa_y=\kappa$ for simplicity, stabilization of $X$ and $Y$ leads to the axino mass:
\bea
m_{\tilde a} \sim \frac{N_M\,\kappa^2}{32\pi^2} \frac{F_Z}{Z_0}\sim 
N_M \left( \kappa \over g \right)^2 m_{\rm soft}
\eea
which can allow a large ratio $m_{\tilde a}/m_{\rm soft} \gg 1$ when $\kappa$ is larger than the standard model gauge coupling $g$ or $N_M$ is large. 
The soft scalar masses for $X$ and $Y$ are generated at two loop level as 
\bea
\tilde m_X^2 = \tilde m_Y^2 \simeq N_M(N_M+2) \left|\frac{ \kappa^2F}{32\pi^2 Z_0}\right|^2.
\eea
They are all positive, so that the saxion can be stabilized without dangerous unstable directions.
Its physical mass is
\bea
m_s^2\sim \left(\frac{ N_M \kappa^2}{32\pi^2} \frac{F_Z}{Z_0}\right)^2 = {\cal O}(m_{\tilde a}^2).
\eea

Note that the dominant SUSY breaking superfields can also be charged 
under the U(1)$_{PQ}$  \cite{Carpenter:2009sw}.
In this kind of model, $R$-symmetry is imposed in the global SUSY limit, and thus there are 
generically light $R$-saxion/axion fields.
One then find the following typical mass spectrum:
\bea
{\rm saxion,\, axino} &:&\, F/Z_0\nonumber\\
R\textrm{-saxion} &:&\, F/4\pi Z_0\nonumber\\
\textrm{MSSM sparticles} &:&\, F/16\pi^2 Z_0\nonumber\\
R\textrm{-axion} &:& \sqrt{Z_0/M_P} F/Z_0 \nonumber\\
{\rm gravitino} &:&  F/M_P.
\eea
with $Z_0\sim f_a$. It also gives a heavy axino/saxion with $m_{\tilde a} \sim m_s \sim 100 m_{\rm soft}$.
The existence of such a light $R$-axion is model-dependent, and might play an important role in cosmology.
We do not study these models in this paper. Related work can be found in \cite{Hamada:2012fr,Hamada:2013xga}.

\section{Phenomenological implications of SUSY axion models}

The `QCD axion,' by definition, has the `anomalous' interaction with gluons:
\begin{equation}
{\cal L}\supset\frac{g_s^2 }{32\pi^2 f_a/N}\,a \,G^{b\mu\nu}\widetilde{G}^b_{\mu\nu},
\label{eq:QCDaxion}
\end{equation}
where $g_s$ is the coupling constant of QCD,
$G^{b\mu\nu}$ is the gluon field strength, and $\widetilde{G}^b_{\mu\nu}$ is its dual.
In SUSY theories, this interaction is supersymmetrized by
\begin{eqnarray}
{\cal L}&\supset&-\frac{\sqrt{2}g_s^2 }{32\pi^2 f_a/N }\int d^2\theta AW^bW^b+{\rm h.c.},
\label{eq:anomaly}
\end{eqnarray}
where $W^b$ is gluon field strength superfield.
It includes interactions of axinos and saxions in addition to  Eq.~(\ref{eq:QCDaxion}).
Note that $f_a$ is related to $v_{PQ}$ as $f_a= \sqrt{2}v_{PQ}$ and $N$ is the domain wall number.


The above Lagrangian is generated after integrating out (heavy) fermions
charged under the anomalous PQ symmetry U(1)$_{PQ}$.
In the linearly realized axion models, U(1)$_{PQ}$ can be realized by coupling
the U(1)$_{PQ}$ breaking singlet superfield $X$  to either color-charged fields (KSVZ)
or the Higgs bilinear operator (DFSZ), or to both:
\begin{equation}
W=\lambda_1 X \Phi \Phi^c+\lambda_2 \frac{X^2}{M_P} H_uH_d ,
\label{eq:genpot}
\end{equation}
where $\Phi+\Phi^c$ is $3+\bar{3}$ under SU(3)$_c$, and $H_{u,d}$ is up (down)-type Higgs multiplet.
This superpotential respects the PQ symmetry with the PQ charge assignment: $(\Phi+\Phi^c,H_u+H_d)=(-1,-2)$.

Note that $N=N_\Phi$ with $N_\Phi$ being the number of $\Phi+\Phi^c$, in the pure KSVZ model ($\lambda_1 \neq 0$ and $\lambda_2=0$), whereas  $N=6$ in DFSZ ($\lambda_1= 0$ and $\lambda_2 \neq 0$).
On the other hand, one has $N=|6-N_\Phi|$ in the hybrid case (KSVZ+DFSZ).
In the following, we will discuss separately phenomenological implication of the KSVZ and DFSZ models
with heavy axino/saxion.

\subsection{KSVZ}

In  the KSVZ superpotential,
\bea\label{KSVZterm}
W = \lambda_1 X \Phi\Phi^c.
\eea
$\Phi+\Phi^c$ can be a larger representation, e.g., $5+\bar{5}$ under SU(5), which includes $3+\bar{3}$ of SU(3)$_c$.
In this case, the axion supermutiplet has the additional anomaly interactions similar to Eq.~(\ref{eq:anomaly}) with
SU(2)$_L$ and U(1)$_Y$ gauge superfields, which has non-trivial  
implications not only to the axion physics but also to the MSSM spectrum.
For the heavy axino scenario under consideration, the ratio $F_X/X_0$ can be considerably 
larger than the gravitino mass as shown in Eq.~(\ref{Fterm}).
Then, the SUSY breaking effect can be mediated to the visible MSSM sector by the gauge interactions of $\Phi+\Phi^c$
and thus sizable soft SUSY breaking terms can be generated.
We call this ``axionic gauge mediation''.
The corresponding soft masses are of order
\bea
\Delta_{\rm a} M_{\rm soft}\equiv \frac{1}{16\pi^2 }\frac{F_X}{X_0}
={\cal O}\Big( \frac{m_{\tilde a}}{ 16\pi^2}\Big).
\eea
If $m_{\tilde a} ={\cal O}(100\,{\rm TeV})$,
and $\Phi+\Phi^c$ are charged under all the SM gauge groups,
the desired soft masses of order TeV can be generated. 
That is, the KSVZ axion model naturally provides gauge mediation with heavy PQ charged
matter fields playing the role of messengers.
In this set-up, one has $m_{\rm soft} \sim m_{\tilde a}/16\pi^2$ and thus
\bea
\frac{m_{3/2}}{m_{\rm soft}} \sim 16\pi^2 {f_a \over M_P} \sim 10^{-4} \,
\left({f_a \over 10^{12} {\rm GeV} } \right) ,
\eea
which realizes again the spectrum of $m_{3/2} \ll m_{\rm soft} \ll m_{\tilde a}$.


\subsection{DFSZ}

An attractive feature of the DFSZ model with the superpotential
\begin{eqnarray}
W=\lambda_2\frac{X^2}{M_P}H_uH_d ,
\label{eq:dfsz}
\end{eqnarray}
is that the $\mu$-term~is generated naturally \cite{KimNilles}:
\begin{equation}
\mu=\lambda_2\frac{X_0^2}{M_P} ={\cal O}\left(\frac{v_{PQ}^2}{M_P}\right).
\end{equation}
Moreover, the non-zero $F$-term generates also the $B\mu$ term in the Higgs scalar potential:
\begin{eqnarray}
{\cal L}&\supset& \int d^2\theta\,
\left(\lambda_2\frac{X_0^2}{M_P}\right)\left(\frac{2 F_X}{X_0}\theta^2\right) H_uH_d=
\frac{2 F_X}{X_0}\mu H_uH_d,
\end{eqnarray}
that is,
\begin{equation}
 B\mu=\frac{2 F_X}{X_0}\mu\sim m_{\ta}\mu\sim m_s\mu.
\end{equation}
On the other hand,
the $\mu$/$B\mu$-terms and $Z$-boson mass are related by the electroweak symmetry breaking condition~\cite{bbhmmt13},
\begin{eqnarray}
\frac{M_Z^2}{2}&=&\frac{(m_{H_d}^2+\Sigma_d^d)-(m_{H_u}^2+\Sigma_u^u)\tan^2\beta}{\tan^2\beta-1}-\mu^2,
\label{eq:ewsb}\\
B\mu&=&\left\{
(m_{H_u}^2+\mu^2+\Sigma_u^u)+(m_{H_d}^2+\mu^2+\Sigma_d^d)
\right\}\sin\beta\cos\beta +\Sigma_u^d,
\label{eq:Bmu}
\end{eqnarray}
where $\Sigma^{u,d}_{u,d}$ is the radiative correction for the Higgs mass parameters.
In the large $\tan\beta$ and decoupling limit, Eq.~(\ref{eq:ewsb}) approximately becomes
\begin{eqnarray}
\frac{M_Z^2}{2}\simeq -\mu^2 -m_{H_u}^2 + m_{H_d}^2/\tan^2\beta,
\end{eqnarray}
neglecting the radiative corrections.
For natural electroweak symmetry breaking, each term in the right-hand side should be of 
order $M_Z^2$.
Thus one needs
\begin{equation}
\mu \sim M_Z \sim {\cal O}(100) \mbox{ GeV}
\end{equation}
which can be achieved if $v_{PQ}\sim10^{10} \, (10^{11})$ GeV for $\lambda_2\sim1 \, (0.01)$.
Moreover, Eq.~(\ref{eq:Bmu}) requires
\begin{eqnarray}
B\mu\simeq (m_{H_u}^2+ m_{H_d}^2)/\tan\beta \simeq m_{H_d}^2/\tan\beta.
\end{eqnarray}
where $|m_{H_u}^2|\ll m_{H_d}^2$ is assumed in the decoupling limit.
Then, the naturalness argument says
\begin{eqnarray}
m_{H_d}^2/\tan^2\beta \simeq B\mu/\tan\beta \lesssim M_Z^2.
\end{eqnarray}
From the relation $B\sim m_{\ta}\sim m_s$ and $\mu \sim M_Z$,
one can put the upper limit for the axino and saxion mass:
\begin{equation}
m_{\ta}\sim m_s \lesssim M_Z\tan\beta.
\end{equation}
Thus, the axino/saxion mass may be required to be below $\sim$ 10 TeV
considering natural electroweak symmetry breaking.

\section{Cosmology with heavy axino/saxion and a gravitino as LSP}

\subsection{Two MSSM benchmark models: SUA and SOA}





%
\begin{table}\centering
\begin{tabular}{lcc}
\hline
 &~~~~SUA~~~~&~~~~SOA~~~~    \\
\hline
$\tan\beta$  & 10 & 10  \\
$M_1$ & 311.3 & 222.2 \\
$M_2$ & 571.5 & 410.6 \\
$\mu $ & 200.0 & 2598 \\
$m_A$ & 1000 & 4284 \\
$m_h$ & 124.8 & 125.0 \\
$m_{\tg}$ & 1793 & 1312 \\
$m_{\tu}$ & 5116 & 3612 \\
$m_{\tst_1}$ & 1226 & 669.0 \\
$m_{\tz_1}$ & 187.7 & 224.1 \\
\hline
$\Omega^{\rm std}_{\tz_1} h^2$ & 0.013 & 6.8 \\
\hline
\end{tabular}
\caption{Masses and parameters in~GeV units for two benchmark points
computed with \Isajet\,7.83 \cite{isasugra} and using $m_t=173.2$ GeV. 
}
\label{tab:bm}
\end{table}

In this section, we will discuss the cosmological implications of heavy axinos and saxions,
concentrating on dark matter properties with the gravitino as the LSP. 
In order to see the effects of next-to-lightest SUSY particle (NLSP) on gravitino production, 
we consider two different benchmark points. 
The first one-- labelled SUA for standard underabundance of NLSP (if it were dark matter)-- 
contains a Higgsino-like neutralino as NLSP. 
The second one-- labelled SOA for standard overabundance-- contains a Bino-like neutralino as NLSP.
In Table~\ref{tab:bm}, some weak scale parameters, sparticle masses and the putative NLSP density are 
shown for these two benchmark points.
An advantage of choosing these two benchmark cases is that the results of the current work with a
gravitino as LSP may be directly compared to previous work with a heavy gravitino but with a neutralino as 
LSP~\cite{bbls14}.


We display here only the weak scale spectra for the SUSY benchmark models  with two different 
cases of a neutralino NLSP.
Although we do not specify any UV-complete models for these scenarios, it is worthwhile providing some 
comments.
Since the gravitino is the LSP, gauge-mediation is a plausible mechanism to produce these 
sparticle mass spectra~\cite{gmsb_review}.
After the discovery of Higgs boson, a number of papers have examined how to obtain a 125 GeV Higgs 
mass in gauge mediation models with relatively light top squarks~\cite{Kang:2012ra,Craig:2012xp,Craig:2013wga,Abdullah:2012tq,Kim:2012vz}.
The Higgsino-like NLSP has also been explored in non-minimal gauge mediation 
models~\cite{Dimopoulos:1996yq,Agashe:1999ct,Agashe:1997kn,Baer:1999tx,Matchev:1999ft,Baer:2000pe,Culbertson:2000am,Cheung:2007es,Meade:2009qv}.
 It is interesting to work out concrete models which reproduce the properties of the above benchmark scenarios. However, it is beyond the scope of this work and thus we leave it for a future task.

\subsection{Thermal and non-thermal gravitino production}

The axino and saxion can be produced efficiently in the early universe by thermal scattering, decay and inverse decays which can alter the standard dark matter property. 
The axino and saxion thermal production has been studied extensively for the 
KSVZ case~\cite{Covi01,Brand04,Strumia10} as well as for the DFSZ case~\cite{Chun11,bci1,bci2}.
Depending on the PQ breaking scale, reheat temperature, and axino mass, it can be either hot, warm or cold dark matter if the axino is sufficiently light~\cite{cckr11}.
In such circumstances, the axion-axino mixed dark matter scenario can also be realized~\cite{Baer:2009ms}.
Along with the axino, the saxion can also play an important role in cosmology and astrophysics~\cite{Kawasaki:2013ae}.
For conventional gravity mediation models with a typical mass spectrum,
 $m_{\ta}\sim m_s\sim m_{3/2}\sim m_{\rm soft}$, the LSP is normally the lightest neutralino, and the decays of the abundant axino and saxion have to be taken into account as they can affect the neutralino relic density. In such a case,
the axion-neutralino mixed dark matter scenario can be realized  either  in the KSVZ model~\cite{ckls,blrs,bls,bbl} or in the DFSZ model~\cite{Chun11,bci2,bbc1,bbc2,bbls14}.

In this work, we address a different possibility: the heavy axino/saxion with light gravitino.
As shown in Sec.~\ref{sec:mass}, the axino and saxion can be much heavier 
than not only the gravitino but also the MSSM sparticles.
In this case, we have two dark matter candidate: the gravitino and the axion.
The axion dark matter is produced from coherent oscillations during the QCD phase transition.
Concerning the gravitino production, there are three different sources in our scenario:
\begin{itemize}

\item {\it thermal production}\\
The gravitinos are produced from the thermal bath via interactions with MSSM particles.
The gravitino thermal density is given by~\cite{Bolz00,Rychkov07}
\begin{equation}
\Omega_{\widetilde{G}}^{\rm TP}h^2=0.21
\left(\frac{m_{\widetilde{g}}}{1\mbox{ TeV}}\right)^2
\left(\frac{1\mbox{ GeV}}{m_{3/2}}\right)\left(\frac{T_R}{10^{8}\mbox{ GeV}}\right)
\label{eq:grav_TP}
\end{equation}
where $T_R$ is the reheat temperature after the primordial inflation, and $m_{\tilde{g}}$ is the 
gluino mass.
As described from this equation, it is possible that a sufficient amount of gravitinos are 
produced from the thermal bath if $T_R$ is large enough.

\item {\it decay of axinos and saxions}\\
The gravitinos are also produced from the decays of axinos and/or saxions.
These decays are extracted from the interaction term~\cite{Cremmer82}
\bea
\frac{1}{2M_P}\partial_\nu (s- i a)  \bar\psi_\mu\gamma^\nu
\gamma^\mu  (1-\gamma_5)\tilde a
+ \mbox{h.c.}
\eea
and the corresponding decay rates are given by~\cite{Chun93}
\bea
\Gamma(\tilde a\to a+\widetilde{G}) &=&\frac{1}{96\pi} \frac{m_{\tilde a}^5}{m_{3/2}^2 M_{ P}^2},\\
\Gamma(s\to \tilde a+\widetilde{G}) &=&\frac{1}{48\pi}
\frac{m_s^5}{ m_{3/2}^2 M_P^2}
\left(1-\frac{m_{\tilde a}^2}{m_s^2}\right)^{4}.\label{eq:sax_tagra}
\eea
In general, the PQ scale is much smaller than the Planck scale,
so thermally produced axinos and saxions are much more abundant than the gravitino.
Hence, this process can be an important source of gravitino production.

\item {\it decay of neutralinos}\\
Neutralino NLSPs are produced from thermal and non-thermal processes and ultimately 
decay into the gravitino LSP.
The gravitino density from neutralino decay is simply determined by the ratio of the 
gravitino mass to neutralino mass and the neutralino density before it decays:
\begin{equation}
\Omega_{\widetilde{G}}^{\widetilde{Z}_1}h^2=\frac{m_{3/2}}{m_{\widetilde{Z}_1}}\Omega_{\widetilde{Z}_1}h^2.
\end{equation}
Therefore, it strongly depends on the neutralino composition of $\tilde Z_1$ which determines the relic density.
An important constraint on the neutralino NLSP decay to the gravitino LSP comes from its impact on  Big Bang Nucleosynthesis (BBN),
which will be discussed in more detail later.
The dominant NLSP decay modes are given by~\cite{wss06}
\begin{eqnarray}
\Gamma(\widetilde{Z}_1\to\widetilde{G}+\gamma)&=&\frac{\left(v_4^{(1)}\cos\theta_W+v_3^{(1)}\sin\theta_W\right)^2}{48\pi m_{3/2}^2M_P^2}m_{\widetilde{Z}}^5,
\label{eq:neut_dec1}\\
\Gamma(\widetilde{Z}_1\to\widetilde{G}+Z)&=&
\frac{2\left(v_4^{(1)}\sin\theta_W-v_3^{(1)}\cos\theta_W\right)^2+\left(v_1^{(1)}\sin\beta-v_2^{(1)}\cos\beta\right)^2}{96\pi m_{3/2}^2M_P^2}\nonumber\\
&&\times m_{\widetilde{Z}_1}^5\left(1-\frac{m_Z^2}{m_{\widetilde{Z}_1}^2}\right)^4,
\label{eq:neut_dec2}\\
\Gamma(\widetilde{Z}_1\to\widetilde{G}+\phi)
&=&\frac{\left|\kappa_{\phi}\right|^2}{16\pi}
m_{\widetilde{Z}_1}^5\left(1-\frac{m_{\phi}^2}{m_{\widetilde{Z}_1}^2}\right)^4,
\label{eq:neut_dec3}
\end{eqnarray}
where $\phi=h,H,A$ and
\begin{eqnarray}
\kappa_h&=&-\frac{(i)^{\theta_1+1}}{\sqrt{6}M_Pm_{3/2}}\left[v_1^{(1)}\cos\alpha+v_2^{(1)}\sin\alpha\right],\\
\kappa_H&=&-\frac{(i)^{\theta_1+1}}{\sqrt{6}M_Pm_{3/2}}\left[-v_1^{(1)}\sin\alpha+v_2^{(1)}\cos\alpha\right],\\
\kappa_A&=&-\frac{(i)^{\theta_1+2}}{\sqrt{6}M_Pm_{3/2}}\left[v_1^{(1)}\cos\beta+v_2^{(1)}\sin\beta\right].
\end{eqnarray}

\end{itemize}
Here, the $v_i^{(1)}$ denote the $i$th component of the lightest neutralino, where
$i=1,2$ corresponds to higgsino, $i=3$ to wino and $i=4$ to bino in the notation of 
Ref. \cite{wss06}.


While the thermal production of gravitinos is simply determined by the gravitino mass and reheat temperature,
the non-thermal productions from the axino/saxion decay and neutralino decay strongly depend on the PQ sector and the MSSM spectrum.
In the following sections, we will examine some specific examples of the MSSM spectrum to 
study these effects separately for the KSVZ, DFSZ and hybrid cases.
 For these analyses, we will assume that the PQ symmetry is already broken during and after inflation, 
so that the Hubble parameter and the reheating temperature are hierarchically smaller than 
the PQ breaking scale.\footnote{If the phase transition of the PQ symmetry occurs after 
the end of inflation, the PQ symmetry breaking scale and the domain wall number are strongly constrained
especially by the axion dark matter abundance produced by strings and domain walls~\cite{Kawasaki:2014sqa}.}

\subsection{KSVZ} 

For the KSVZ axion model, Eq.~(\ref{eq:anomaly}) is the only relevant interaction with the MSSM sector.
Having only dimension-five interactions, the thermal yields of the axion/saxion are proportional to the reheat temperature $T_R$~\cite{Brand04,Strumia10,Graf12}:
\begin{eqnarray}
Y_{\tilde{a}}^{\rm TP}&=&0.9\times10^{-5}g_s^6\ln\left(\frac{3}{g_s}\right)
\left(\frac{10^{12}\mbox{ GeV}}{f_a}\right)^2
\left(\frac{T_R}{10^{8}\mbox{ GeV}}\right),
\label{eq:axn_TP}\\
Y_{s}^{\rm TP}&=&1.3\times10^{-5}g_s^6
\ln\left(\frac{1.01}{g_s}\right)
\left(\frac{10^{12}\mbox{ GeV}}{f_a}\right)^2
\left(\frac{T_R}{10^{8}\mbox{ GeV}}\right).
\label{eq:sax_TP}
\end{eqnarray}
For saxions, coherent oscillations can also lead to a large yield given by
\begin{eqnarray}
Y_{s}^{\rm CO}&=&1.9\times10^{-5}
\left(\frac{\mbox{min}\left[T_R,T_s\right]}{10^8\mbox{ GeV}}\right)
\left(\frac{f_a}{10^{12}\mbox{ GeV}}\right)^2
\left(\frac{\mbox{GeV}}{m_s}\right)
\label{eq:sax_CO}
\end{eqnarray}
where $T_s$ is the temperature at which the saxion field starts to oscillate:  $3H(T_s)=m_s$.

Here we assumed that the initial displacement of the saxion field is $f_a$, {\it i.e.} $s_0=f_a$.
Taking an initial value of $s_0$ as $f_a$ is a natural choice 
since generic supergravity effects provide additional Hubble-induced
mass terms for the saxion field. 
With a modified scalar potential, the saxion becomes heavy 
with a mass of ${\cal O}(H)$ for $H\gg m_s$, and stays in its modified vacuum value during inflation. 
As $H$ decreases, the saxion field follows the instantaneous minimum,
and begins to oscillate when $H \sim m_s$.  
At this moment,  
the displacement from its present value would just be ${\cal O}(f_a)$. 
For example, in models like $W \sim (XY - f_0^2)S$, 
the additional Hubble induced SUSY breaking terms just change the ratio between $X_0$ and $Y_0$ 
while fixing $X_0Y_0 =f_0^2$. Without fine-tuning we easily expect 
$\delta X_0\sim \delta Y_0= {\cal O}(f_0) = {\cal O}(f_a)$, 
which implies a saxion amplitude of ${\cal O}(f_a)$.
Meanwhile, in the model of Eq. (\ref{eq:model}), the situation becomes more interesting
because for $m_{3/2}\ll H\ll f_a$, the saxion mass becomes
${\cal O}( (H^2 f_a)^{1/3})$ which is much greater than $H$
for generic Hubble-induced SUSY breaking terms. 
In such a case, the saxion is strongly captured near its minimum, and 
adiabatically moves to its effective vacuum value even for $H \lesssim m_{3/2}$. 
Thus, here the oscillating amplitude is very small. 
This kind of phenomena is studied in the context of the moduli problem~\cite{Linde:1996cx}. 
Here we do not calculate detailed oscillation amplitudes, but instead
in our forthcoming $T_R$ bounds, 
we consider a case with $s_0 = 0.01\, f_a$ as an example corresponding to the model (\ref{eq:model}).

The produced axinos and saxions decay mainly into gluons and gluinos through the interactions in
Eq.~(\ref{eq:anomaly}).  For the saxion decay, we note that from Eq.~(\ref{eq:FA_term}) and 
Eq.~(\ref{eq:axino_mass}) 
$F_A$ also depends on $s$ and its coefficient is proportional to the axino mass. Thus,
 in addition to the standard interactions, the additional saxion-gluino-gluino interaction 
can be obtained as 
\bea
{\cal L}\supset \frac{\sqrt{2} g_s^2}{32\pi^2 f_a/N}\, F_A\, \tilde g^{\alpha b} \tilde g^b_\alpha  + \mbox{h.c.} 
\to -\frac{g_s^2 m_{\tilde a}}{16\pi^2 f_a/N}\, s\, \tilde g^{\alpha b} \tilde g^b_\alpha  + \mbox{h.c.}.
\eea
Then the partial decay widths are given by
\begin{eqnarray}
\Gamma\left(\ta\to\tilde{g}+g\right)&=&
\frac{\alpha_s^2}{16\pi^3f_a^2}m_{\ta}^3
\left(1-\frac{m_{\tilde{g}}^2}{m_{\ta}^2}\right)^3,\\
\Gamma\left(s\to g+g\right)&=&
\frac{\alpha_s^2m_s^3}{32\pi^3f_a^2},\\
\Gamma\left(s\to \tilde{g}+\tilde{g}\right)&=&
\frac{\alpha_s^2(m_{\tilde{g}}+m_{\ta})^2m_s}{8\pi^3f_a^2}
\left(1-\frac{4m_{\tilde{g}}^2}{m_s^2}\right)^{3/2} .
\end{eqnarray}
If $\ta\to\tilde{g}g$ and/or $s\to\tilde{g}\tilde{g}$ are not kinematically allowed, we 
should also consider the decays via the electromagnetic interactions similar to Eq.~(\ref{eq:anomaly}), 
which leads to
\begin{eqnarray}
\Gamma(\tilde{a}\to\widetilde{Z}_i+\gamma)&=&\frac{\left(\alpha_YC_{aYY}\cos\theta_Wv_4^{(i)}\right)^2}{128\pi^3(f_a/N)^2}m_{\tilde{a}}^3\biggl(1-\frac{m_{\widetilde{Z}_i}^2}{m_{\tilde{a}}^2}\biggr)^3
\\
\Gamma(\tilde{a}\to\widetilde{Z}_i+Z)&=&\frac{\left(\alpha_YC_{aYY}\sin\theta_Wv_4^{(i)}\right)^2}{128\pi^3(f_a/N)^2}m_{\tilde{a}}^3\lambda^{1/2}\biggl(1,\frac{m_{\widetilde{Z}_i}^2}{m_{\tilde{a}}^2},\frac{m_Z^2}{m_{\tilde{a}}^2}\biggr)\nonumber\\
&&\cdot\biggl\{
\biggl(1-\frac{m_{\widetilde{Z}_i}^2}{m_{\tilde{a}}^2}\biggr)^2+3\frac{m_{\widetilde{Z}_i}m_Z^2}{m_{\tilde{a}}^3}-\frac{m_Z^2}{2m_{\tilde{a}}^2}\biggl(1+\frac{m_{\tilde{Z}_i}^2}{m_{\tilde{a}}^2}+\frac{m_Z^2}{m_{\tilde{a}}^2}\biggr)
\biggr\},
\end{eqnarray}
\begin{eqnarray}
\Gamma(s\to Z+Z)&=&
\frac{\left(\alpha_YC_{aYY}\sin^2\theta_W\right)^2}{256\pi^3 (f_a/N)^2}\nonumber\\
&&\cdot m_s^3
\left(
1-\frac{4m_Z^2}{m_s^2}
\right)^{1/2}
\left(
1-\frac{4m_Z^2}{m_s^2}+\frac{6m_Z^4}{m_s^4}
\right),\\
\Gamma(s\to \gamma+\gamma)&=&
\frac{\left(\alpha_YC_{aYY}\cos^2\theta_W\right)^2}{256\pi^3 (f_a/N)^2}m_s^3,\\
\Gamma(s\to Z+\gamma)&=&
\frac{(\alpha_YC_{aYY})^2\sin^2\theta_W\cos^2\theta_W}{128\pi^2(f_a/N)^2}
m_s^3\left(1-\frac{m_Z^2}{m_s^2}\right)^4,\\
\Gamma(s\to\widetilde{Z}_i+\widetilde{Z}_j)&=&
\frac{\left(\alpha_YC_{aYY}v_4^{(i)}v_4^{(j)}\right)^2}{128\pi^3(f_a/N)^2}\lambda^{1/2}\biggl(1,\frac{m_{\widetilde{Z}_i}^2}{m_s^2},\frac{m_{\widetilde{Z}_j}^2}{m_s^2}\biggr)\left(1-\frac12\delta_{ij}\right)\nonumber\\
&&\cdot m_s(m_{\widetilde{Z}_i}+m_{\widetilde{Z}_j}+2m_{\ta})^2\biggl[
1-\biggl(\frac{m_{\widetilde{Z}_i}+m_{\widetilde{Z}_j}}{m_s}\biggr)^2
\biggr],
\end{eqnarray}
where $C_{aYY}=(0,2/3,8/3)$ for the heavy quark charges $e_{\Phi}=(0,-1/3,+2/3)$.

For the saxion, there are additional decay modes into axions and axinos from the effective
Lagrangian for the axion supermultiplet:
\bea\label{eq:Lag_ax_sup}
{\cal L} \supset \left(1+ \frac{2}{f_a}s \right)
\left\{ \frac{\xi}{2} (\partial^\mu a )(\partial_\mu a) +
\frac{\xi'}{2}(\partial^\mu s)(\partial_\mu s)
+ \frac{i\xi''}{2}\bar{\tilde a }\partial\hspace{-2.0mm}/\tilde a\right\} ,
\eea
from which one finds 
\begin{eqnarray}\label{eq:saxion_dec}
\Gamma(s\to a+a)&=&\frac{\xi^2m_s^3}{32\pi f_a^2},\label{eq:sax_aa}\\
\Gamma(s\to\ta+\ta)&=&\frac{\xi''^2m_{\ta}^2m_s}{4\pi f_a^2}\left(1-\frac{4m_{\ta}^2}{m_s^2}\right)^{3/2},
\end{eqnarray}
where $\xi$, $\xi'$ and $\xi''$ are the model-dependent constants 
determined by the effective interactions in Eq.~(\ref{eq:axion1}).
In general, $\xi$, $\xi'$ and $\xi''$ are not the same, but if $F_A=0$ and $Z_3^F=0$, $\xi=\xi'=\xi''$ as in Ref.~\cite{Chun95}.
In this work, we assume $\xi=\xi''$ in the following analyses for simplicity.


The heavy axinos decay into lighter particles and thus affect the density of those light spices.
The amount of non-thermal gravitinos from axino decay is determined by the axino density and its decay branching fraction:
\begin{equation}
\Omega^{\ta}_{\widetilde{G}}h^2=2.8\times10^8\left(\frac{m_{3/2}}{\rm GeV}\right)
BR(\ta\to a+\widetilde{G})~Y_{\ta}.
\label{eq:grav_axn_dec}
\end{equation}
Comparing the major decay modes of the axino, one gets
\begin{eqnarray}
\frac{\Gamma(\ta\to a+\widetilde{G})}{\Gamma(\tilde{a}\to g+\tilde{g})} 
&=&\frac{\pi^2}{6\alpha_s^2}\frac{f_a^2m_{\ta}^2}{m_{3/2}^2M_P^2}\nonumber\\
&\sim& 10^2\left(\frac{F_X}{F_{\rm tot}}\right)^2.
\label{eq:br_ta_gra}
\end{eqnarray}
where $F_{\rm tot}\equiv\sqrt{3m_{3/2}^2M_P^2}$.
It is interesting to note that the branching fraction is determined by
the ratio of $F$-terms of the PQ sector  and the dominant SUSY breaking sector.
Due to the factor of  ${\cal O}(10^2)$, $\Gamma(\ta\to a+\widetilde{G})$ can be sizable or even 
the dominant decay mode for large $f_a^2m_{\ta}^2$.

\begin{figure}
\begin{center}
\includegraphics[width=7.5cm]{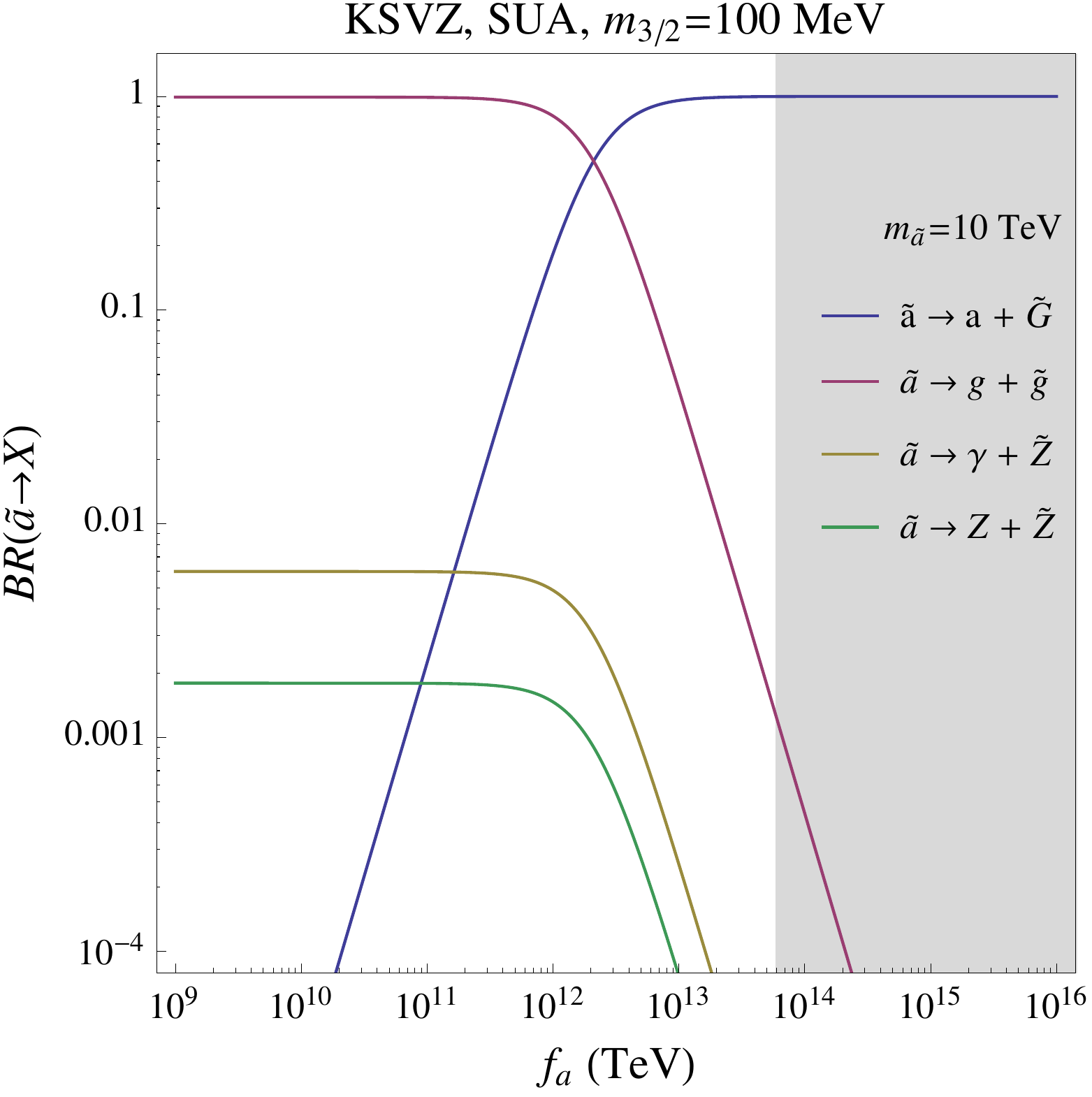}
\includegraphics[width=7.5cm]{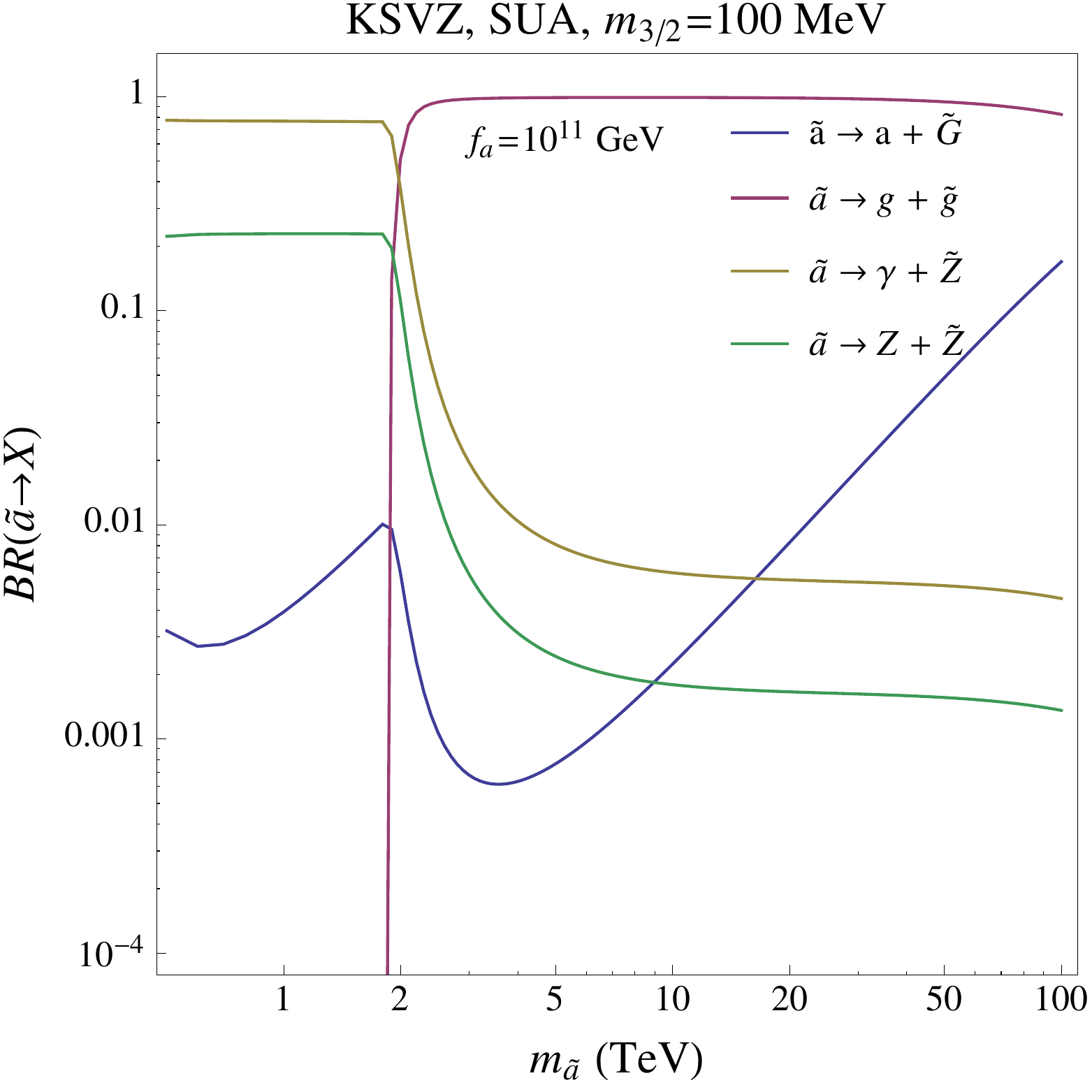}
\end{center}
\caption{ $a)$ Axino branching fractions versus $f_a$ for $m_{\ta}=10$ TeV (left)
and $b)$ versus $m_{\ta}$ for $f_a=10^{11}$ GeV (right) in the KSVZ model. 
The sparticle mass spectrum is taken from the SUA benchmark point of Ref.~\cite{bbls14}.
\label{fig:br_axn}}
\end{figure}

As shown in Fig.~\ref{fig:br_axn}{\it a}), the mode $\ta\to a+\widetilde{G}$  becomes dominant
for $f_a>10^{12}$ GeV.
However, for $f_a\gtrsim 5\times10^{13}$ GeV, a 10 TeV axino mass violates 
the self-consistency condition (\ref{eq:axn_con}) and thus the corresponding region is shaded out.
In the case of $f_a\lesssim10^{12}$ GeV,  the gravitino density from axino decay takes a simple form:
\begin{eqnarray}
\Omega_{\widetilde{G}}^{\ta}h^2\simeq 0.05\,\left(\frac{m_{\ta}}{10\mbox{ TeV}}\right)^2
\left(\frac{100\mbox{ MeV}}{m_{3/2}}\right)\left(\frac{T_R}{10^5\mbox{ GeV}}\right).
\label{eq:axn_dec_grav_den}
\end{eqnarray}
This relation is valid for $m_{\ta}\gtrsim10$ TeV.
For smaller axino mass, the branching fraction of $\ta\to a+\widetilde{G}$ can be enhanced by kinematic suppression of $\ta\to g+\tilde{g}$ modes or small weak gauge coupling of $\ta\to Z/\gamma +\widetilde{Z}$ mode.
For $m_{\ta}\lesssim 2$ TeV, the branching fraction to a gravitino final state is 
an order of magnitude larger than that for $m_{\ta}\gtrsim 2$ TeV as shown in Fig.~\ref{fig:br_axn}{\it b}).
For $f_a\gtrsim 10^{12}$ GeV, $BR(\ta\to a+\widetilde{G})\simeq1$, so the gravitino density from axino decay becomes
\begin{equation}
\Omega_{\widetilde{G}}^{\ta}h^2\simeq 0.003
\left(\frac{m_{3/2}}{100\mbox{ MeV}}\right)
\left(\frac{10^{13}\mbox{ GeV}}{f_a}\right)^2
\left(\frac{T_R}{10^5\mbox{ GeV}}\right).
\end{equation}

Similar to axino decay, the saxion can also decay into gravitinos if allowed kinematically.
The gravitino production from saxion decay can be determined by the branching fraction to 
the gravitino final state.
For $m_s\gtrsim 10$ TeV, the saxion dominantly decays into an axion pair if $\xi \sim 1$.
From Eqs. (\ref{eq:sax_tagra}) and (\ref{eq:sax_aa}), we can estimate the decay fraction:
\begin{eqnarray}
\frac{\Gamma(s\to \ta+\widetilde{G})}{\Gamma(s\to a+a)} 
&=&\frac{2}{3\xi^2}\frac{m_s^2f_a^2}{m_{3/2}^2M_P^2}\nonumber\\
&\sim&{\cal O}(1)\left(\frac{F_X}{F_{\rm tot}}\right)^2.
\label{eq:br_sax_gra}
\end{eqnarray}
Comparing this with Eq. (\ref{eq:br_ta_gra}), we easily see that the 
saxion contribution to gravitino production is always smaller than the 
axino contribution if we consider just the thermally-produced axinos and saxions.
In the case of saxions, however, the coherent oscillation of the saxion field for the 
large $f_a$ region becomes the dominant source of saxion production.
We find that the density of gravitinos from saxion CO is given by
\begin{equation}
\Omega_{\widetilde{G}}^{s}h^2\simeq 0.01
\left(\frac{{\rm min}[T_R,T_s]}{10^5\mbox{ GeV}}\right)
\left(\frac{f_a}{10^{13}\mbox{ GeV}}\right)^4
\left(\frac{m_s}{20\mbox{ TeV}}\right)
\left(\frac{m_{3/2}}{100\mbox{ MeV}}\right),
\label{eq:saxCO_grav}
\end{equation}
and thus it may become the dominant gravitino production mode.

The last component of gravitino production is neutralino decay.
Neutralinos are produced by thermal scattering and decays of the particles which are in thermal 
equilibrium. They are also produced by out-of-equilibrium decays of heavy particles.
If the axino and saxion decay before neutralino freeze-out, the decay products are thermalized so that 
axino and saxion decays do not affect the neutralino density.
If the axino and saxion decay after neutralino freeze-out, on the other hand, they produce a huge amount of neutralinos, and the neutralinos quickly re-annihilate into a smaller density.
The neutralino yield after re-annihilation is approximately determined by the annihilation rate 
at the axino/saxion decay temperature as
\begin{equation}
Y_{\widetilde{Z}_1}(T_D^{\ta,s})\simeq\frac{H(T^{\ta,s}_D)}{\langle \sigma v\rangle (T^{\ta,s}_D) s(T^{\ta,s}_D)}
\end{equation}
where $H(T^{\ta,s}_D)$, $\langle \sigma v\rangle (T^{\ta,s}_D)$ and $s(T^{\ta,s}_D)$ are respectively the Hubble parameter, annihilation rate, and entropy density at the decay temperature of axino (saxion), $T=T_D^{\ta,s}$.

\begin{figure}
\begin{center}
\includegraphics[width=7.5cm]{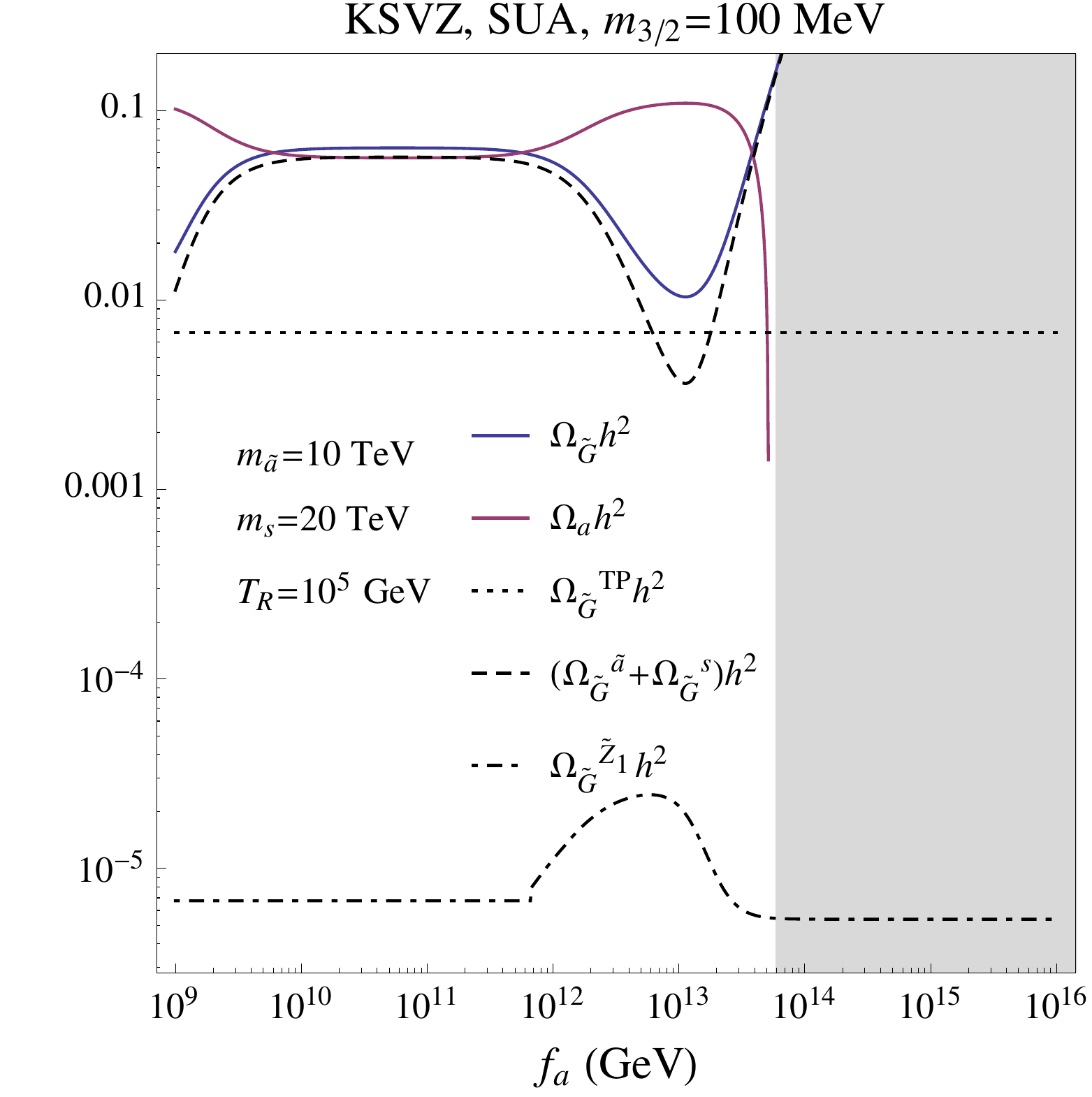}
\includegraphics[width=7.5cm]{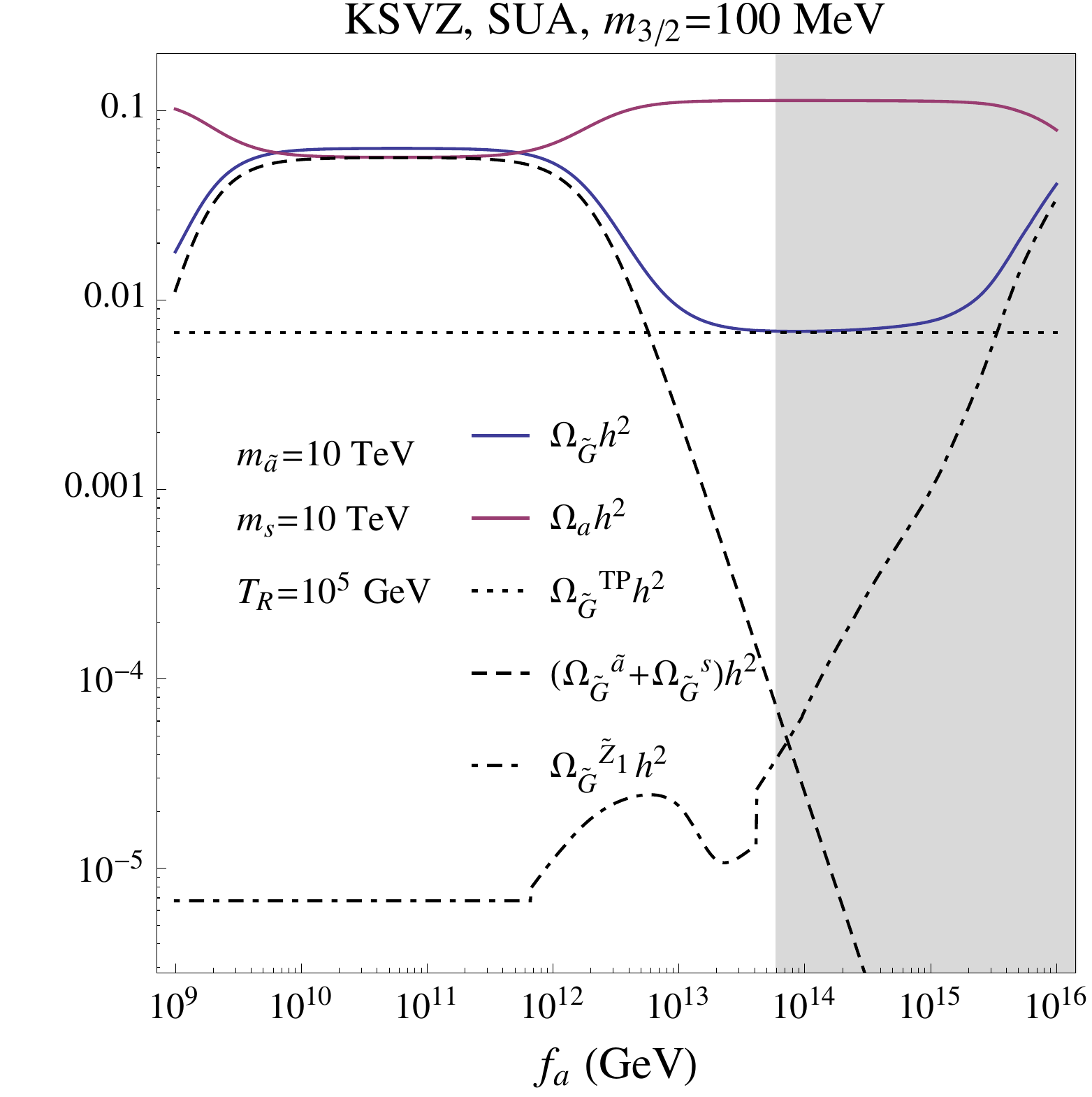}
\end{center}
\caption{Relic abundance of gravitinos from various sources
versus $f_a$ with {\it a}) $m_s= 2 m_{\ta}$ (left)  and {\it b}) $m_s=m_{\ta}$ (right) 
for the SUA benchmark point in the KSVZ model.
\label{fig:oh2_grav_KSVZ_SUA}}
\end{figure}

In Fig.~\ref{fig:oh2_grav_KSVZ_SUA}, we show examples of the gravitino relic 
density as a function of $f_a$ for $a)$ $m_s=2m_{\ta}$ where the saxion can produce gravitinos 
and $b)$  $m_s=m_{\ta}$  which does not allow the saxion decay into gravitinos.
We set $T_R=10^5$ GeV so that the thermal production of gravitinos is not their dominant source.
In both cases, the density of gravitinos from neutralino decay is determined by 
the standard neutralino freeze-out density since the axino and saxion decay 
temperatures are larger than neutralino freeze-out temperature 
($T_{\rm fr}=7$ GeV for SUA).
For $f_a\lesssim10^{12}$ GeV, therefore, the gravitino density is mostly determined from 
axino production and decay.
It is worth noting that for $f_a\lesssim10^{10}$ GeV the axino and saxion thermal production 
is determined by their in-equilibrium values. 
For $10^{12}$ GeV$\lesssim f_a\lesssim 10^{13}$ GeV, the gravitino density becomes smaller since the axino thermal production is getting smaller due to suppression from the increasing 
PQ scale and $BR(\ta\to a+\widetilde{G})$ approaches unity.
Thus, in this region, axions from CO can be the dominant dark matter component.
For $f_a\gtrsim10^{13}$ GeV, two plots show different features.
In the case {\it a}) where $s\to\ta+\widetilde{G}$ is open, gravitino production 
from saxion decay becomes the dominant source of dark matter production since the saxion CO increases as $f_a$ increases.
Therefore, the gravitino density is drastically increasing and becomes larger 
than the overclosure limit when $f_a\gtrsim4\times10^{13}$ GeV.
In the case {\it b}) where the mode $s\to\ta+\widetilde{G}$ is forbidden, 
saxion decay does not contribute to gravitino production.
The increasing neutralino density, which is due to the late decay of saxion CO, is the 
dominant source of gravitino production for $f_a\gtrsim 10^{15}$ GeV.
This region is, however, theoretically inconsistent as argued in Eq.~(\ref{eq:axn_con}).

\begin{figure}
\begin{center}
\includegraphics[width=7.5cm]{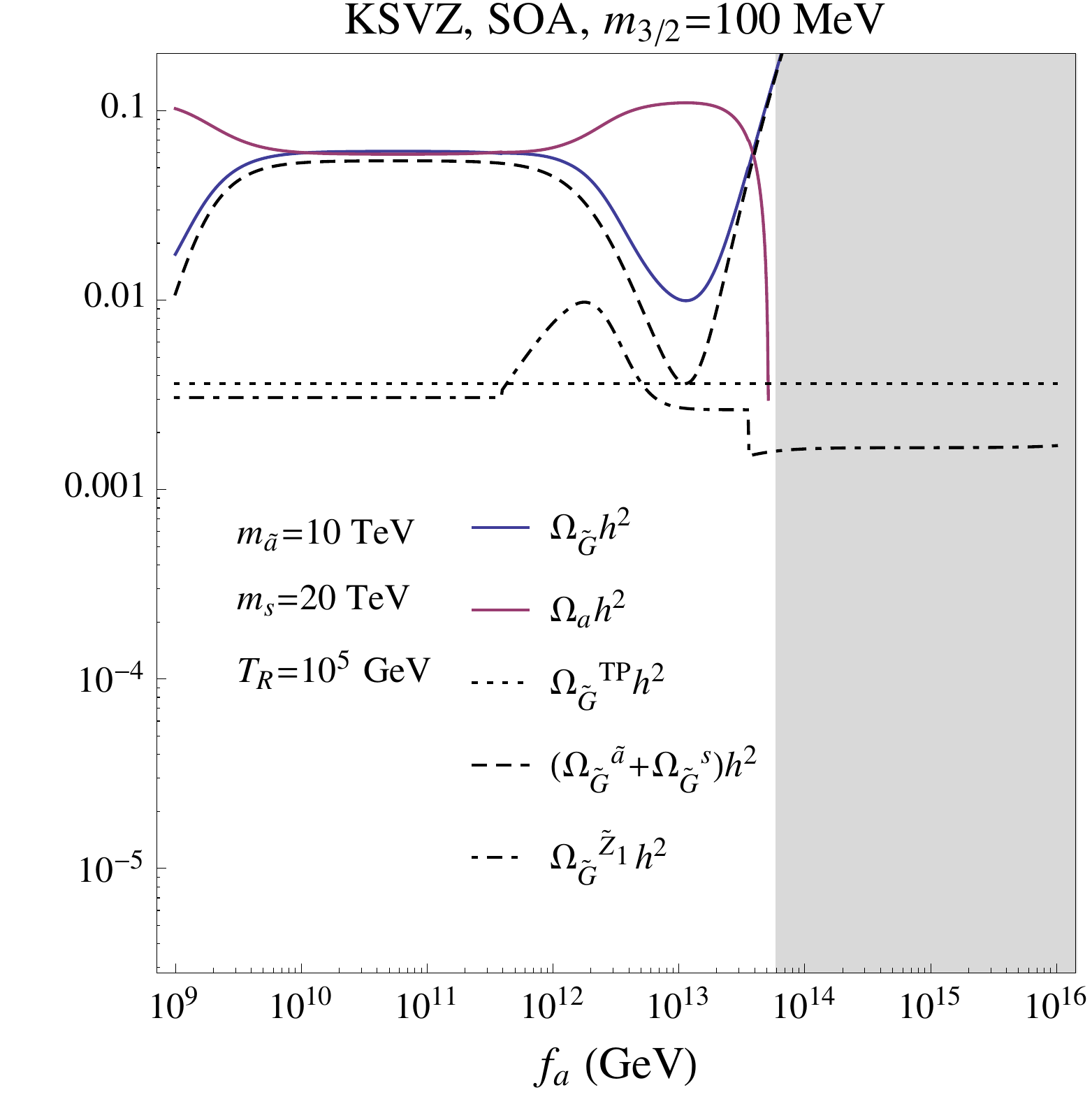}
\includegraphics[width=7.5cm]{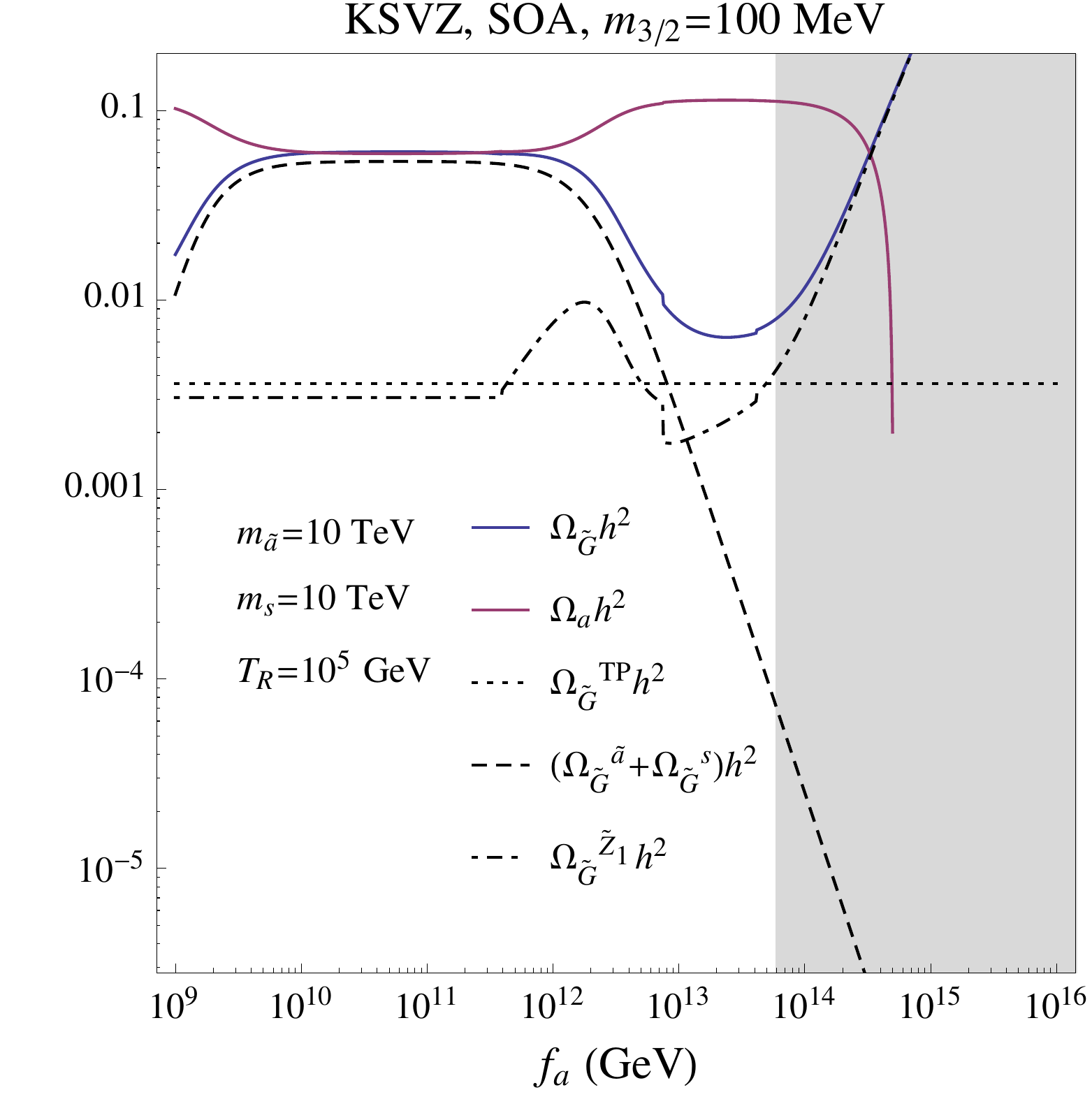}
\end{center}
\caption{Relic abundance of gravitinos from various sources versus $f_a$ 
with {\it a}) $m_s=2 m_{\ta} $ (left)  and {\it b}) $m_s= m_{\ta}$ (right) 
for the SOA benchmark point in the KSVZ model.
\label{fig:oh2_grav_KSVZ_SOA}}
\end{figure}

\medskip

Let us now discuss the SOA benchmark scenario with the Bino-like lightest neutralino for 
a comparison of the SUA benchmark point in which the lightest neutralino is Higgsino-like.
In Fig.~\ref{fig:oh2_grav_KSVZ_SOA}, the gravitino density plots for $a)$ $m_s=2m_{\ta}$  and $b)$ $m_s=m_{\ta}$  are shown.
Most of the physical characteristics are similar to the SUA case except that the pair annihilation cross-section of Bino-like neutralino is much smaller than Higgsino-like neutralinos 
and thus the neutralino density  tends to be larger than the SUA case which is shown
Fig.~\ref{fig:oh2_grav_KSVZ_SOA}{\it b}) for $f_a\gtrsim10^{15}$ GeV.

\begin{figure}
\begin{center}
\includegraphics[width=7.5cm]{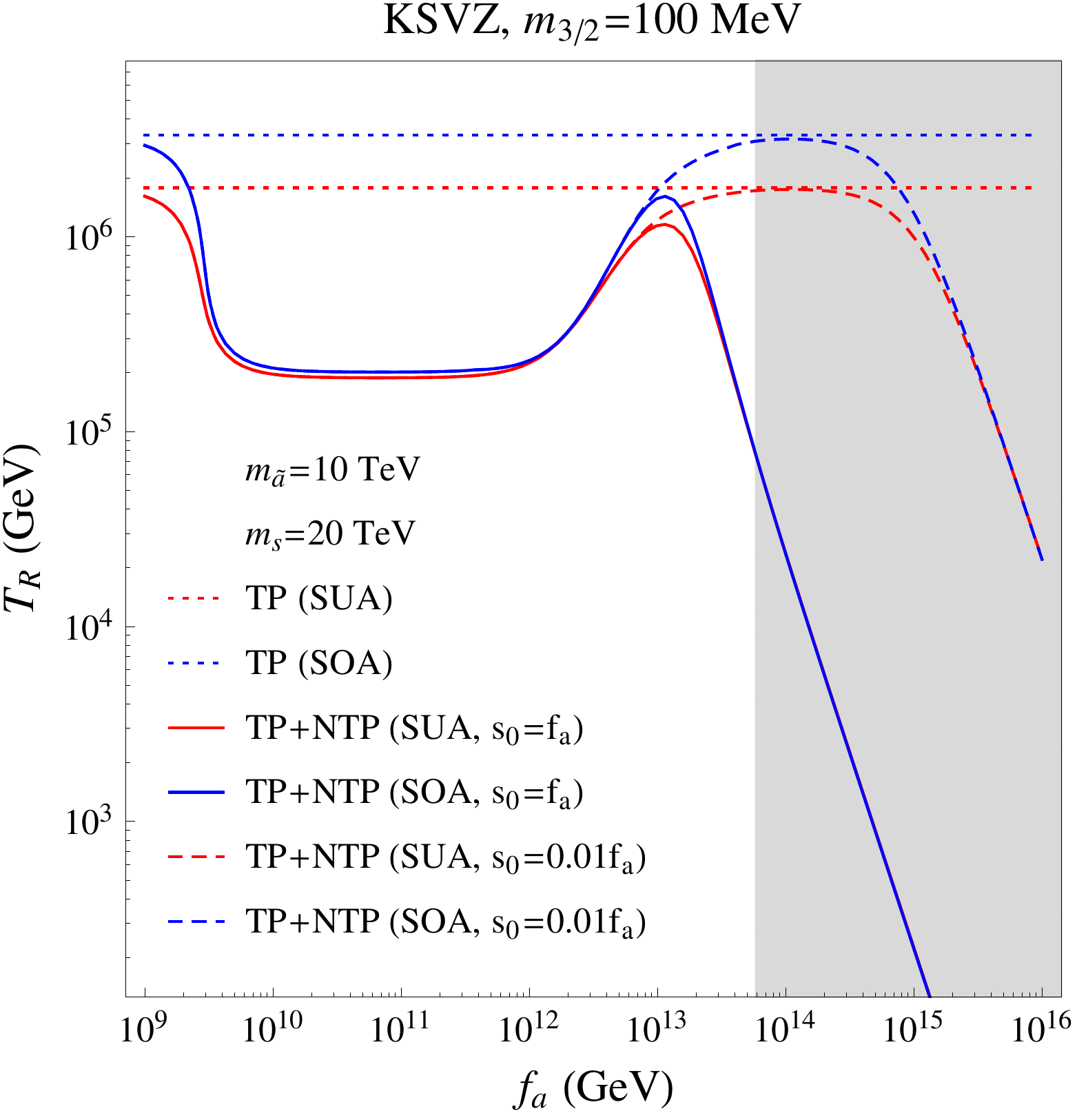}
\includegraphics[width=7.5cm]{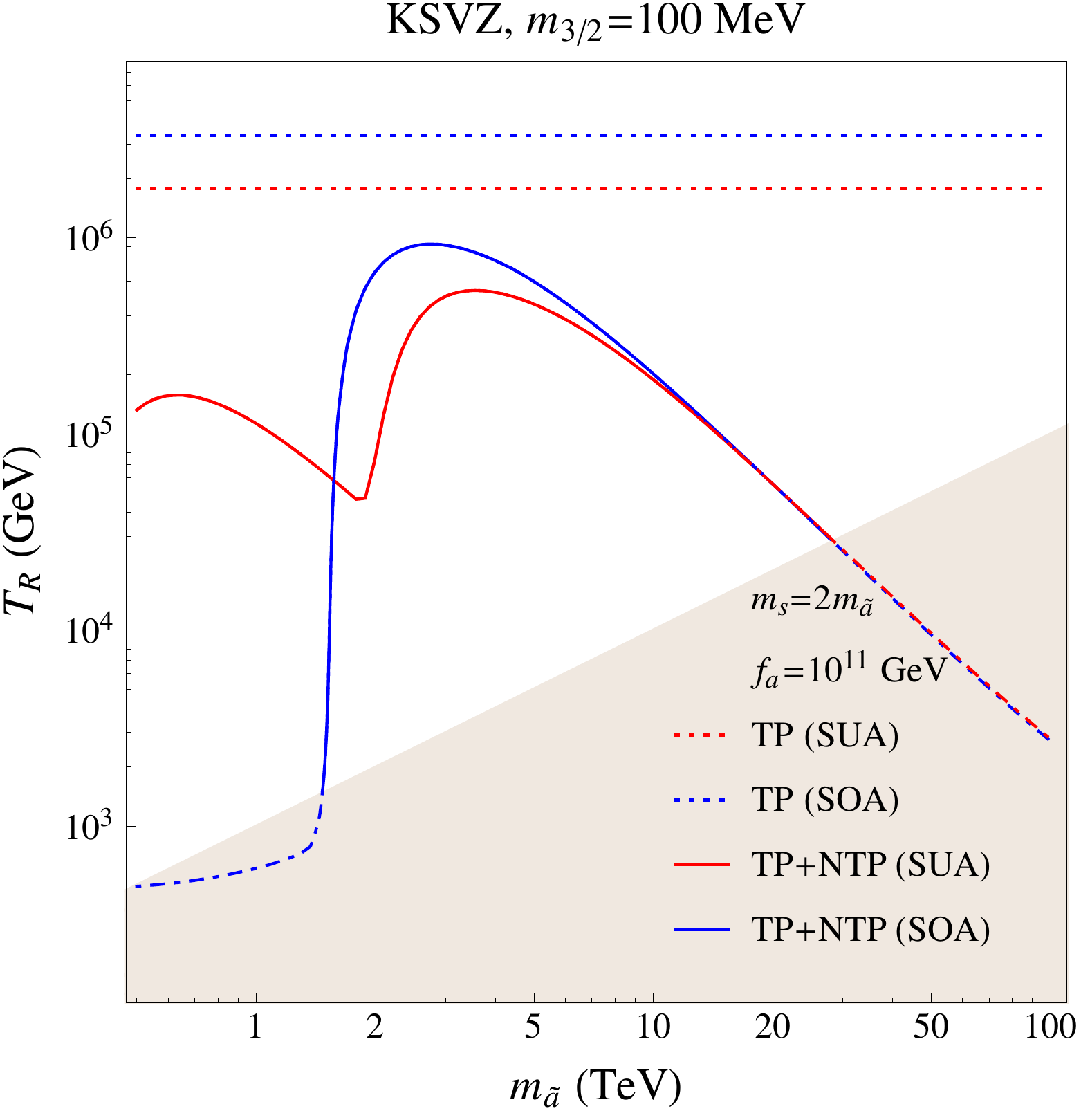}
\end{center}
\caption{The upper bound of $T_R$ calculated from 
thermal and non-thermal production of gravitinos as a function of $a)$ $f_a$   (left)
and $b)$ $m_{\ta}$ (right) for SUA (red) and SOA (blue) in the KSVZ model. 
The region above the curves is disallowed by overproduction of gravitinos.
\label{fig:tr_bound_KSVZ}}
\end{figure}

From the previous discussions that show sizable non-thermal gravitino production 
from the axino/saxion decay depending non-trivially on the axion scale $f_a$ and the axino/saxion mass, 
one can see that the thermal gravitino production has to be suppressed appropriately by putting  
an upper limit on the reheat temperature $T_R$ as a function of $f_a$ and the axino/saxion mass.
Fig.~\ref{fig:tr_bound_KSVZ} shows the $T_R$ bound in terms of  $a)$ $f_a$  and $b)$ $m_{\ta}$ 
assuming $m_s = 2 m_{\ta}$ for both cases. 
Recall that the major source of the non-thermal gravitino density is from 
axino decay for $f_a\lesssim 10^{13}$ GeV and from saxion decay for $f_a\gtrsim10^{13}$ GeV.
The upper limit of the reheat temperature is reduced by an order of magnitude 
for $10^{10}$ GeV $\lesssim f_a\lesssim10^{12}$ GeV where the gravitino production from 
axino decay is maximized. 
For $f_a \gtrsim 10^{13}$ GeV, the $T_R$ bound starts to decrease again as the 
coherent saxion production becomes sizable.
 Meanwhile, as discussed in the beginning of this subsection, $s_0$ can be much smaller than $f_a$.
In this case, saxion CO contribution to the gravitino production is suppressed so that it becomes dominant for larger $f_a\gtrsim 10^{15}$ GeV.  The upper bound on $T_R$ for $s_0=0.01f_a$ (dashed curves) is also shown in Fig.~\ref{fig:tr_bound_KSVZ}{\it a}). 
The right panel of Fig.~\ref{fig:tr_bound_KSVZ} for a fixed $f_a=10^{11}$ GeV 
shows that the $T_R$ bound tends to decrease
as $m_{\ta}$ increases. This can be understood from the fact that $F_X\sim f_a m_{\ta}$  becomes larger and thus 
enhances the branching fraction of the axino decay into gravitinos for larger $m_{\ta}$.
If the axino mass becomes larger than 30 TeV, the $T_R$ bound becomes smaller than the axino mass and thus the formula Eq.~(\ref{eq:axn_TP}) is invalidated. 
In this paper we do not consider the region $T_R < m_{\ta}$ or $m_s$ which is  
shaded out in Fig.~\ref{fig:tr_bound_KSVZ}{\it b}). 
The continuing dot-dashed line shows the bound if Eq.~(\ref{eq:axn_TP}) were still valid.
It is expected that the upper bound of $T_R$ is in the shaded region above the dot-dashed line.
Meanwhile, a clear difference between the SOA and SUA cases can be 
seen in the region of small $m_{\ta}\lesssim2$ TeV.
In this region, the axino and saxion tend to decay after the neutralino freeze-out, and thus there appears an overall enhancement in the neutralino density producing a lot of gravitinos.  
As a consequence, the $T_R$ bound becomes much stronger.

\begin{figure}
\begin{center}
\includegraphics[width=7.5cm]{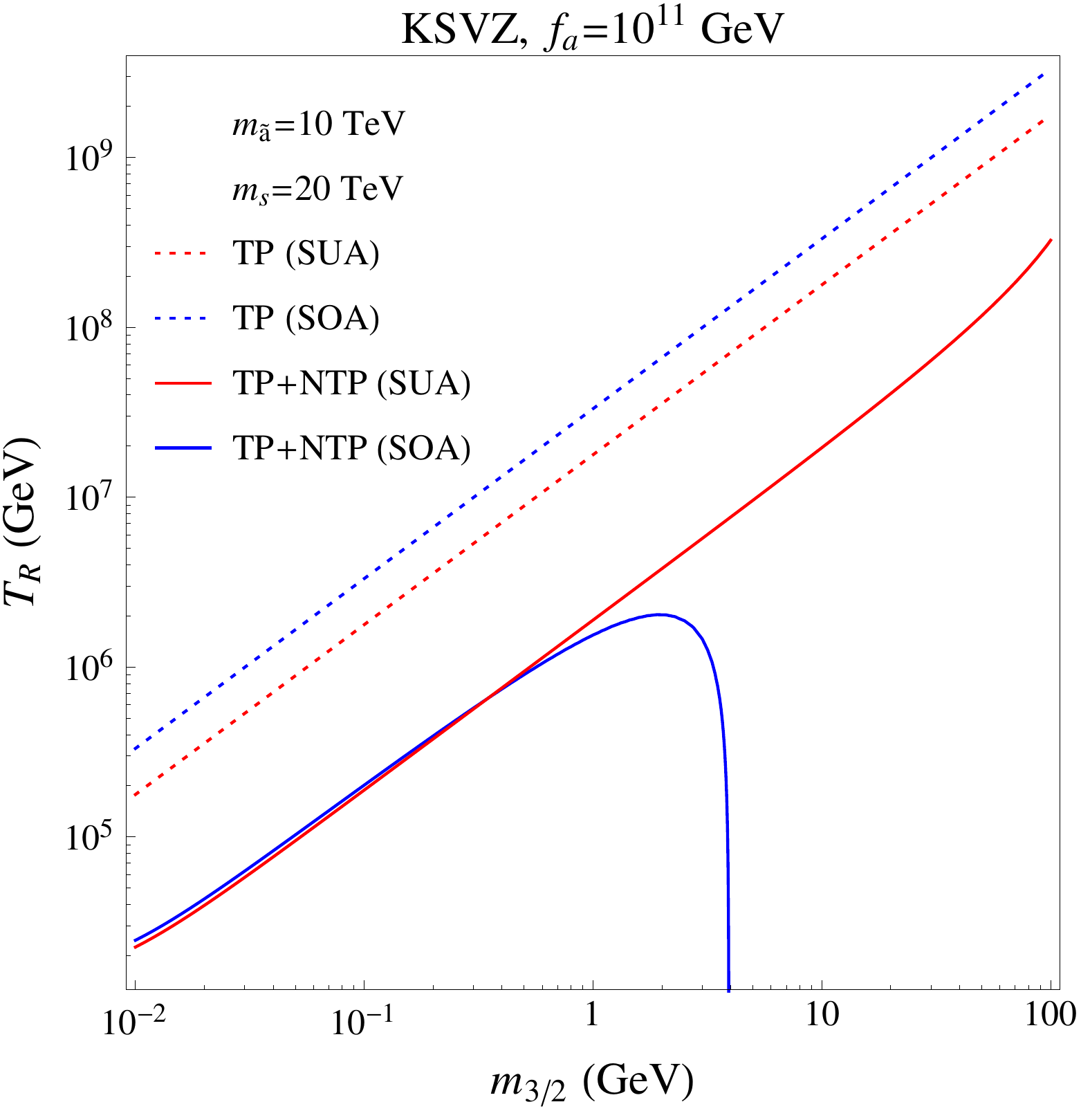}
\end{center}
\caption{The upper bound of $T_R$ calculated from 
thermal and non-thermal production of gravitinos as a function of $m_{3/2}$ for SUA (red) and SOA (blue) in the KSVZ model. 
The region above the curves is disallowed by overproduction of gravitinos.
\label{fig:tr_bound_KSVZ_gra}}
\end{figure}

For different values of $m_{3/2}$ shown is 
the upper bound of $T_R$ in Fig.~\ref{fig:tr_bound_KSVZ_gra} with fixed $f_a=10^{11}$ GeV and $m_{\ta}=m_s/2=10$ TeV.
For $f_a=10^{11}$ GeV, the gravitino density is mostly determined by the non-thermal production from axino and saxion decay as  discussed in the previous paragraphs.
Therefore, the upper bound of $T_R$ is determined by Eq.~(\ref{eq:axn_dec_grav_den}), which is consisitent with the plots. 
For SOA case, however, the upper bound of $T_R$ steeply drops around $m_{3/2}=4$ GeV above which
the gravitino density from neutralino decay exeeds the overclosure limit so that this region is not allowed independently of $T_R$.

\subsection{DFSZ} 

In the DFSZ case, the $\mu$ term operator (\ref{eq:dfsz}) determines the 
axino/saxion interactions which can be written as 
\begin{equation}
{\cal L}_{\rm DFSZ}=\int d^2\theta~ \mu\left(1+B\theta^2\right)H_uH_d e^{-2A/v_{PQ}}+\mbox{h.c.}
\end{equation}
where $B$ is the soft SUSY breaking term in the Higgs sector.
From the above interaction, one finds the axino/saxion population from thermal production given by
\begin{eqnarray}
Y_{\tilde{a}}^{\rm TP}&=&10^{-7}\zeta_{\tilde{a}}
\left(\frac{\mu}{100\mbox{ GeV}}\right)^2
\left(\frac{10^{11}\mbox{ GeV}}{f_a}\right)^2
\left(\frac{\mbox{TeV}}{M_{\rm th}}\right),
\label{eq:axn_TP2}
\\
Y_{s}^{\rm TP}&=&10^{-7}\zeta_{s}
\left(\frac{\mu}{100\mbox{ GeV}}\right)^2
\left(\frac{10^{11}\mbox{ GeV}}{f_a}\right)^2
\left(\frac{\mbox{TeV}}{M_{\rm th}}\right),
\label{eq:sax_TP2}
\end{eqnarray}
where $M_{\rm th}$ is a threshold scale of the process, which can be either Higgsino mass, Higgs mass or axino/saxion mass
and $\zeta_{\tilde{a},s}$ are ${\cal O}(1)$ constants determined by the mass spectrum.
Notice that the thermal yields are independent of the reheat temperature as 
axino/saxion interactions are of the Yukawa type with the coupling  $\mu/v_{PQ}$.

The decays of the DFSZ axino and saxion can be complicated as many channels can open 
due to their mixing with neutralinos and Higgses~\cite{bbc2}.
For the heavy axino ($m_{\ta}\gg \mu$), however, the decay width of the axino is simply given by
\begin{equation}
\Gamma(\ta\to \mbox{Higgsinos}) \simeq \frac{2}{\pi}\left(\frac{\mu}{f_a}\right)^2m_{\ta}.
\end{equation}
On the other hand, the decay width for the gravitino final state is the same as in the KSVZ case,  and thus 
one finds
\begin{eqnarray}
\frac{\Gamma(\ta\to a+\widetilde{G})}{\Gamma(\tilde{a}\to \mbox{Higgsinos})} 
&=&\frac{1}{192}\left(\frac{m_{\ta}}{\mu}\right)^2
\frac{m_{\ta}^2f_a^2}{m_{3/2}^2M_P^2}\nonumber\\
&\sim& 10^{-2} \left(\frac{m_{\ta}}{\mu}\right)^2\left(\frac{F_X}{F_{\rm tot}}\right)^2.
\end{eqnarray}
Since  $F_X  < F_{\rm tot}$, the decay mode $\ta\to a+\widetilde{G}$ is typically subdominant unless 
the axino mass is exceptionally larger than $\mu$.
From the relation, (\ref{eq:grav_axn_dec}) and (\ref{eq:axn_TP2}), we find the gravitino density from the axino decay:
\begin{equation}
\Omega_{\widetilde{G}}^{\ta}h^2\sim 3\times 10^{-4}\zeta_{\ta}
\left(\frac{m_{\ta}}{10\mbox{ TeV}}\right)^3\left(\frac{100\mbox{ MeV}}{m_{3/2}}\right),
\label{eq:oh2_gra_axn_DFSZ}
\end{equation}
where $M_{\rm th}=m_{\ta}$ is taken.

For the saxion decay,  in the case $s\to \ta+\tG$ is open and 
the mode $s\to a+a$ is dominant in the SUA benchmark (with $\xi\sim 1$),
then the saxion branching fraction into the gravitino final state is similar to the KSVZ case. 
From the Eqs. (\ref{eq:br_sax_gra}) and (\ref{eq:sax_TP2}), 
the relic density of gravitinos produced from saxion decay is then
\begin{equation}
\Omega^s_{\widetilde{G}}h^2\sim 3\times10^{-6}\zeta_s\left(\frac{\mu}{100\mbox{ GeV}}\right)^2
\left(\frac{m_s}{10\mbox{ TeV}}\right)\left(\frac{100\mbox{ MeV}}{m_{3/2}}\right),
\label{eq:oh2_gra_sax_DFSZ}
\end{equation}
where we take $M_{\rm th}=m_s$.
As shown in Eq.~(\ref{eq:saxCO_grav}), 
the saxion CO can make a sizable contribution to gravitino production for $f_a\gtrsim10^{13}$ GeV.
In the SOA benchmark, on the other hand, dominant saxion decay can be into Higgses and gauge bosons due to the large $\mu$ term (for $m_s<m_A$),
\begin{equation}
\Gamma(s\to \mbox{Higgses/gauge bosons})\simeq\frac{2}{\pi}\left(\frac{\mu^4}{f_a^2}\right)\frac{1}{m_s}.
\end{equation}
The gravitino from the saxion decay is given by
\begin{equation}
\Omega_{\widetilde{G}}^{s}h^2\sim 8\times10^{-8}\zeta_s
\left(\frac{m_s}{\rm TeV}\right)^5
\left(\frac{2.5 \mbox{ TeV}}{\mu}\right)^2
\left(\frac{100 \mbox{ MeV}}{m_{3/2}}\right).
\end{equation}
For $m_s\gtrsim 5$ TeV, $s\to a+a$ becomes dominant also in the SOA benchmark, so the gravitino density from the saxion decay is the same as Eq.~(\ref{eq:oh2_gra_sax_DFSZ}).

The gravitino production from neutralino decay tends to be similar to the KSVZ case. 
A notable diffence is that  the tree-level Yukawa type coupling $\mu/ v_{PQ}$  makes 
the decays of axino and saxion more rapid (for a given value of $m_{\ta}$ or $m_s$) 
than in KSVZ case so that neutralino decay tends to occur at earlier times: before 
neutralino freeze-out or before onset of BBN.
Then the resulting neutralino density tends to be less sensitive to 
the chosen parameters than in the KSVZ case.

%
\begin{figure}[t]
\begin{center}
\includegraphics[width=7.5cm]{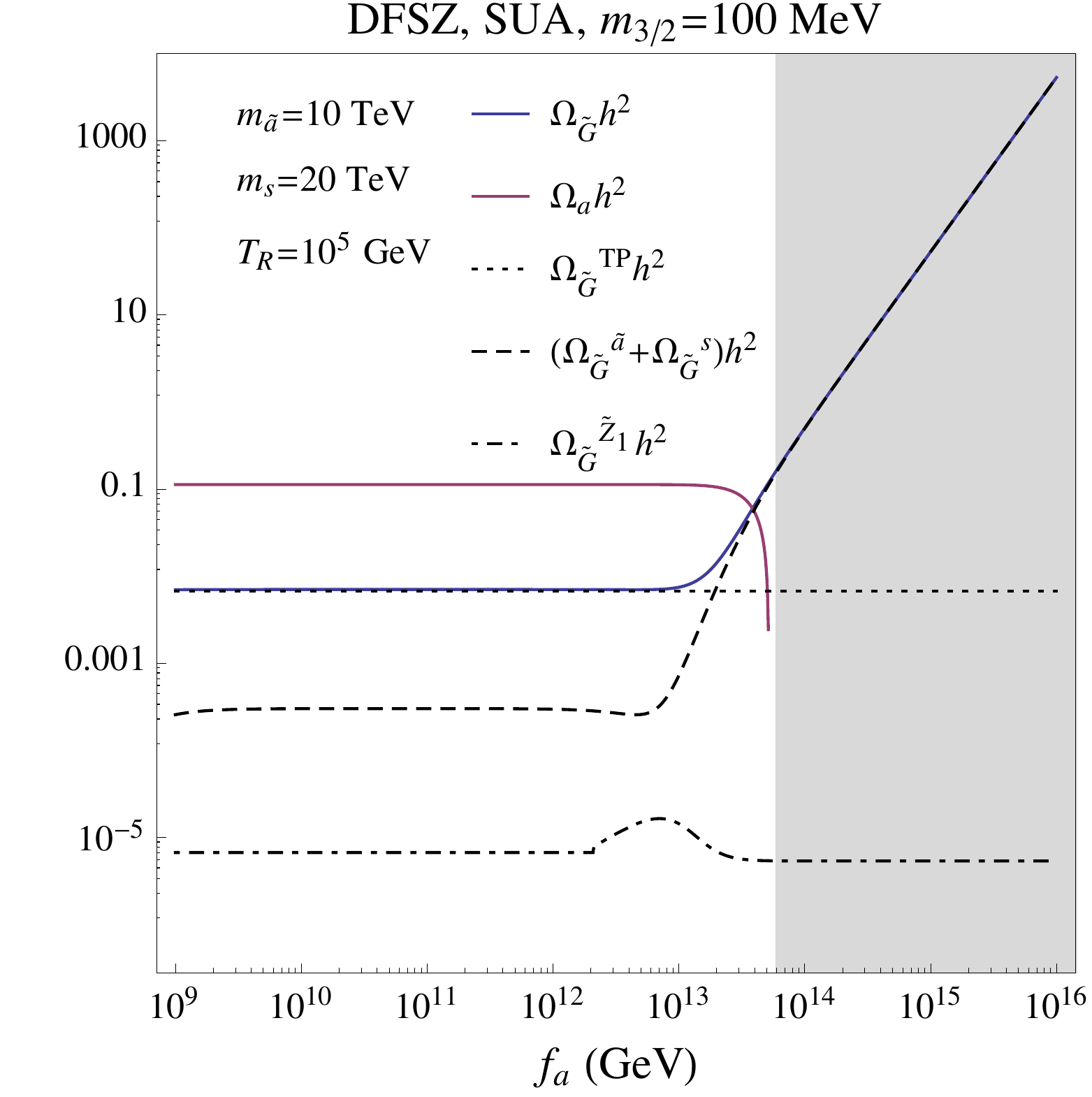}
\includegraphics[width=7.5cm]{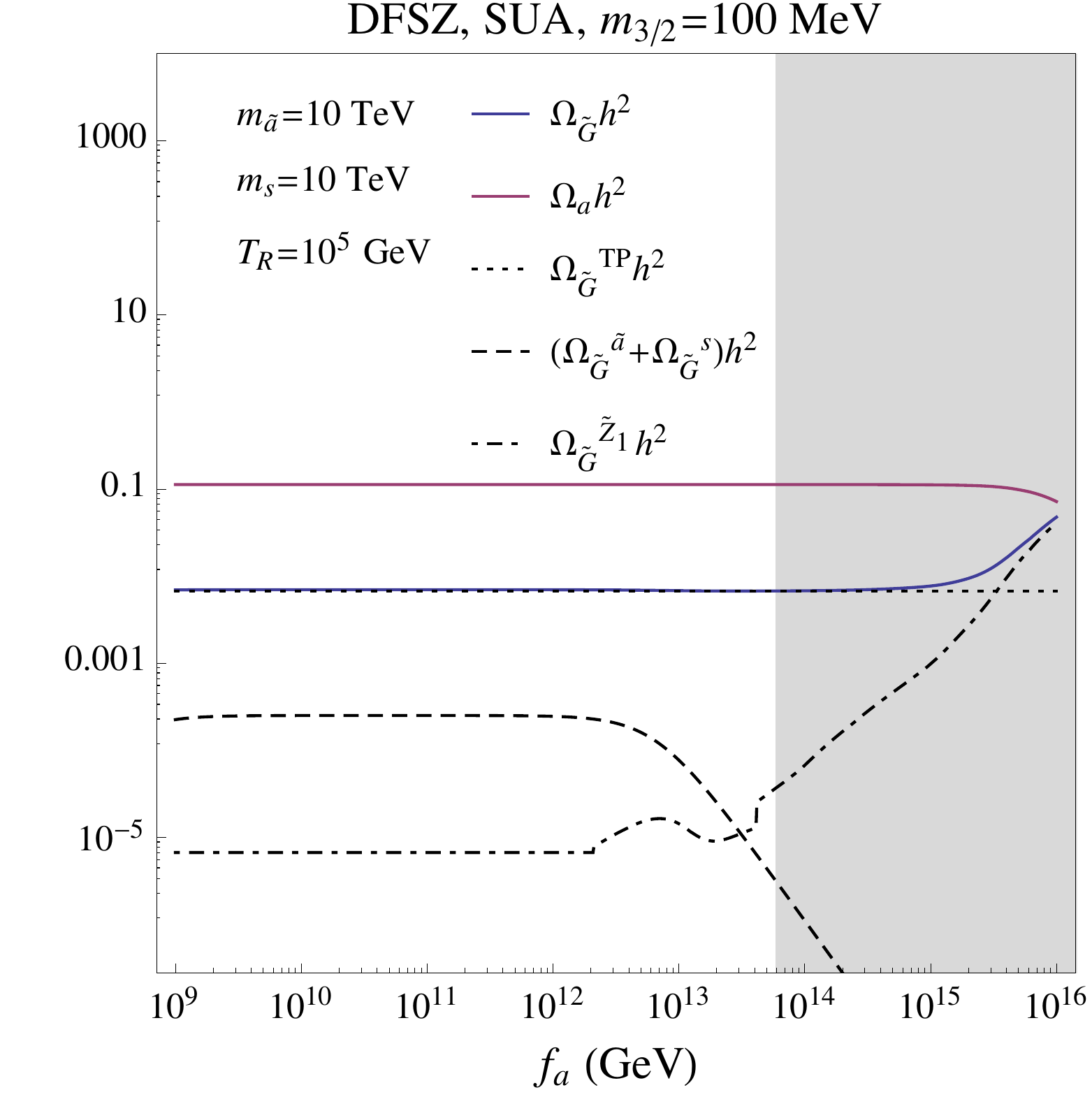}
\end{center}
\caption{Relic abundance of gravitinos from various sources
versus $f_a$ with {\it a}) $m_s= 2 m_{\ta}$  (left) and {\it b}) $m_s=m_{\ta}$ (right) 
for the SUA benchmark point in the DFSZ model.
\label{fig:oh2_grav_DFSZ_SUA}}
\end{figure}
In Fig.~\ref{fig:oh2_grav_DFSZ_SUA}, we show the gravitino density for  the SUA benchmark point 
with $a)$ $m_s=2m_{\ta}$ and $b)$ $m_s=m_{\ta}$.
Notice that the shape of the gravitino density plot is similar to the KSVZ case.
For $f_a\lesssim 10^{12}$ GeV, the axino and saxion decay before the neutralino freeze-out so that the gravitino production 
from the neutralino decay is given by ${\Omega^{\rm std}}_{\widetilde{Z}_1}h^2(m_{3/2}/m_{\widetilde{Z}_1})$.
For $10^{12}$ GeV$\lesssim f_a\lesssim10^{13}$ GeV, the late decays of axino and saxion enhance the neutralino density 
but it is still negligible for the gravitino production.
For $f_a\gtrsim10^{13}$ GeV, saxion CO becomes the dominant source for gravitino production 
if the saxion decay $s\to \ta+\tG$ is open. 
In the case of $m_s=m_{\ta}$, on the contrary,  the saxion decay to neutralinos 
augments the neutralino density which becomes the dominant gravitino source but it occurs only in the
theoretically inconsistent region.

\begin{figure}[t]
\begin{center}
\includegraphics[width=7.5cm]{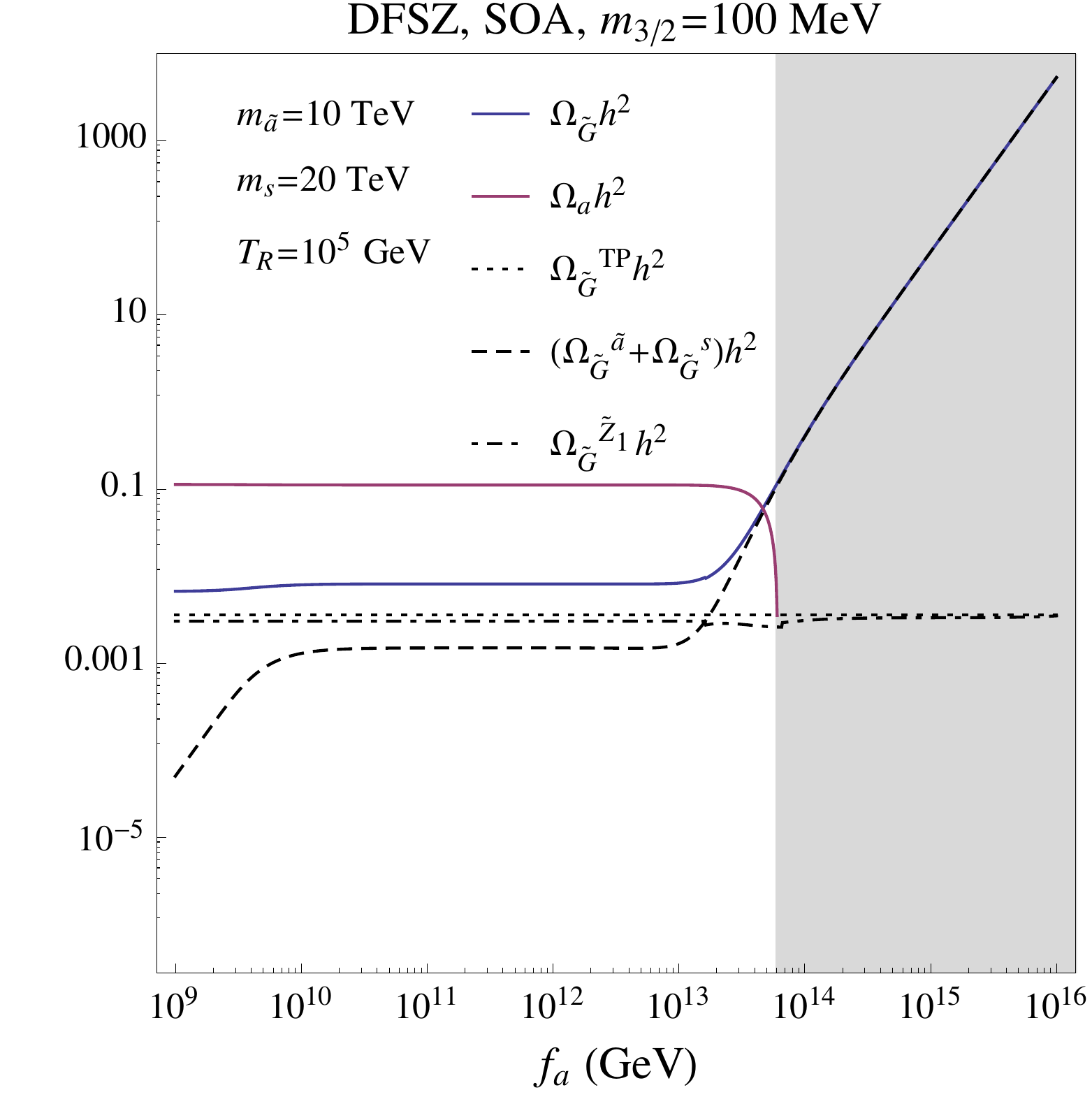}
\includegraphics[width=7.5cm]{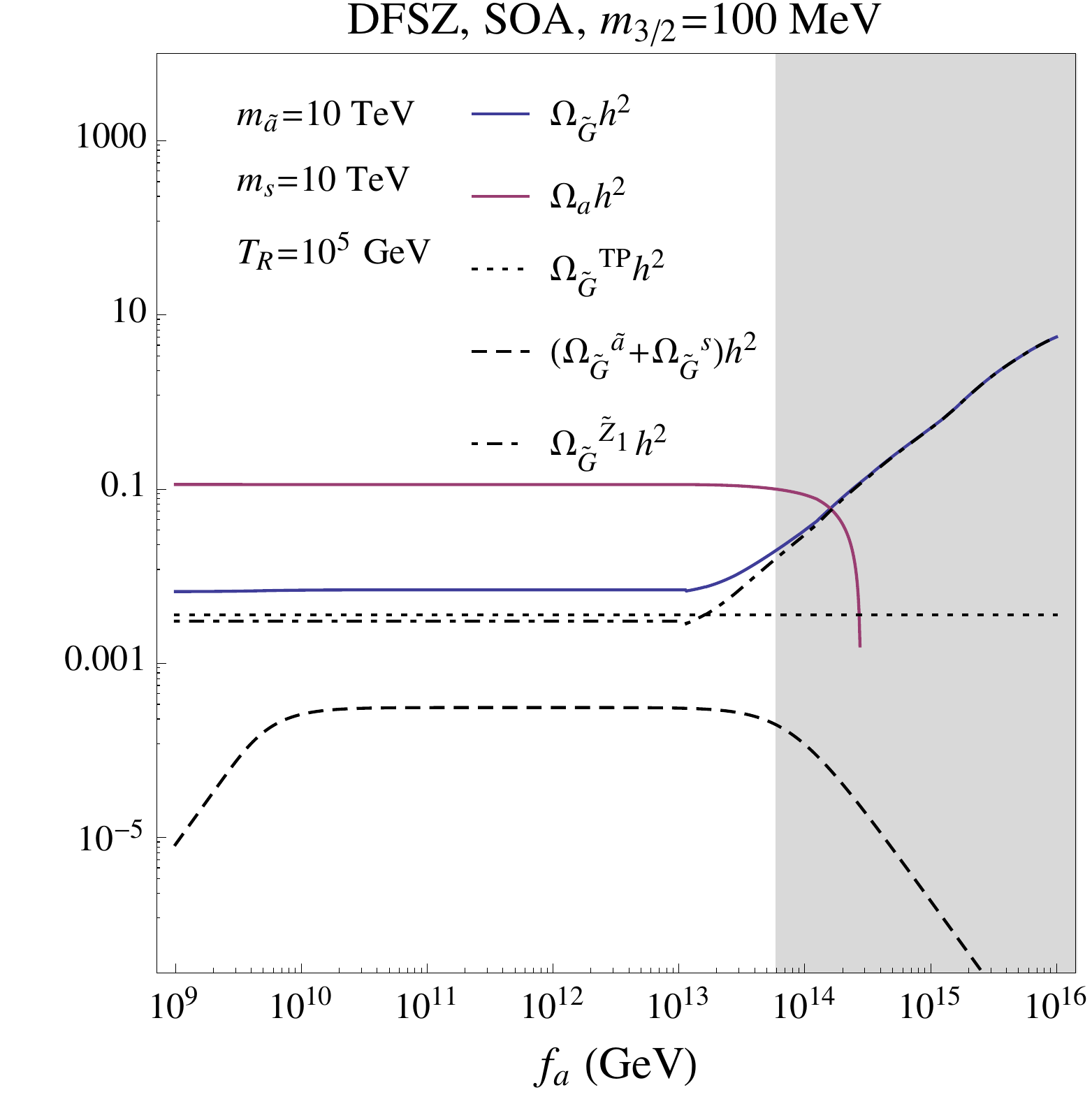}
\end{center}
\caption{Relic abundance of gravitinos from various sources
versus $f_a$ with {\it a}) $m_s= 2 m_{\ta}$ (left)  and {\it b}) $m_s=m_{\ta}$ (right) 
for the SOA benchmark point in the DFSZ model.
\label{fig:oh2_grav_DFSZ_SOA}}
\end{figure}
In Fig.~\ref{fig:oh2_grav_DFSZ_SOA}, we show gravitino density plots for the SOA benchmark.
The large $\mu$-term ($\mu\sim 2.5$ TeV) in this case makes the axino and saxion interactions more efficient so that they tend to decay earlier than neutralino freeze-out even for large $f_a$ up to about $10^{13}$ GeV.
Similar to the SUA case, the saxion CO becomes the dominant source for the gravitino production in the case of $m_s=2m_{\ta}$.
In the case of $m_s=m_{\ta}$, the augmented neutralino density enhances the gravitino density for $f_a\gtrsim10^{13}$ GeV.

\begin{figure}
\begin{center}
\includegraphics[width=7.5cm]{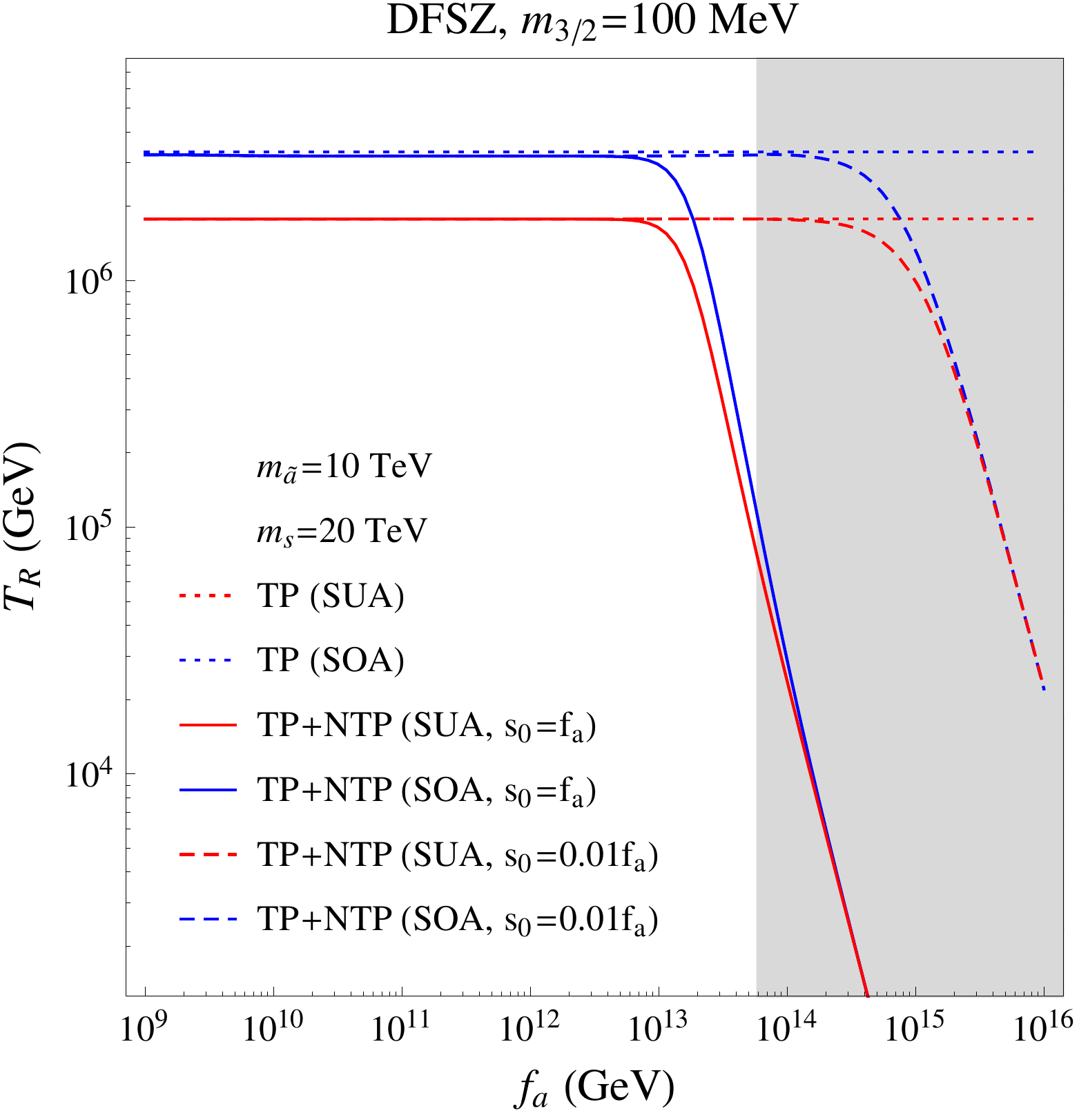}
\includegraphics[width=7.5cm]{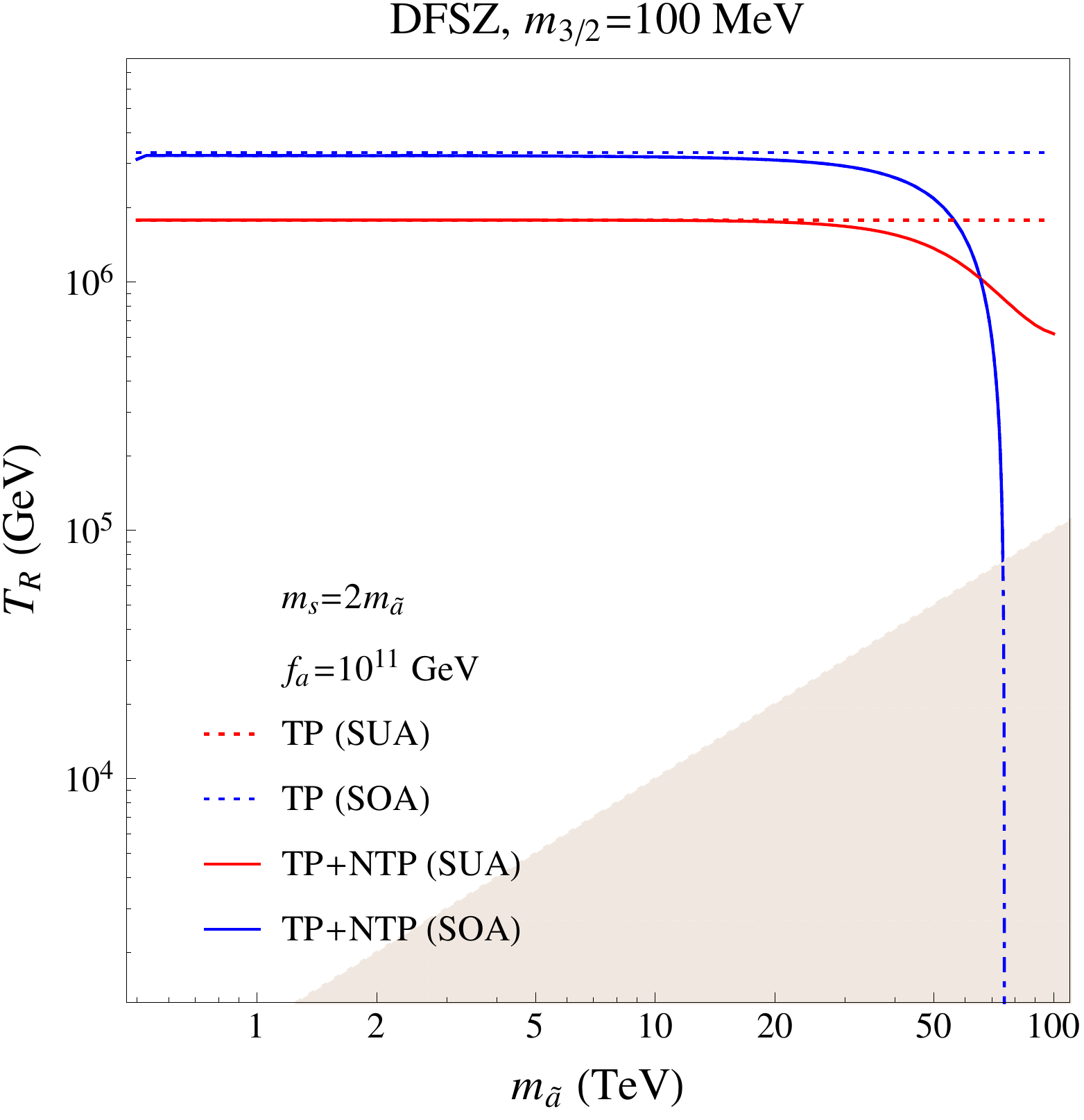}
\end{center}
\caption{The upper bound of $T_R$ calculated from 
thermal and non-thermal production of gravitinos as a function of $a)$ $f_a$ (left) 
and $b)$  $m_{\ta}$ (right) for SUA (red) and SOA (blue) in the DFSZ model. 
The region above the curves is disallowed by overproduction of gravitinos.
\label{fig:tr_bound_DFSZ}}
\end{figure}

The axino and saxion thermal production rates in SUSY DFSZ do not depend on $T_R$ 
while gravitino production from axino and saxion decays is almost independent of $f_a$ 
as shown in Eqs.~(\ref{eq:oh2_gra_axn_DFSZ}) and (\ref{eq:oh2_gra_sax_DFSZ}) 
(as far as  the branching ratios of the axino/saxion decay to the gravitino is less than one).
Thus, the upper limit of $T_R$ is mostly determined by the thermal gravitino production and is independent of $f_a$ for $f_a\lesssim10^{13}$ GeV.
For $f_a\gtrsim10^{13}$ GeV, the dominant gravitino source is the saxion CO which is proportional to $T_R$ and also to $f_a$, 
and thus the $T_R$ bound is steeply decreasing as shown in the left panels of Fig.~\ref{fig:tr_bound_DFSZ}.
 As discussed in the KSVZ case, small $s_0$ makes the saxion CO less effective for the gravitino production.
For $s_0=0.01f_a$, the saxion CO becomes important for larger $f_a\gtrsim 10^{15}$~GeV.
Meanwhile, for lower axino mass, the gravitino abundance from axino and saxion decays is 
negligible so that the $T_R$ bound is determined by the thermal component of gravitino production.
For $m_{\ta}\gtrsim 20$ TeV, however, the axino decay to gravitino becomes sizable, and thus  
the $T_R$ bound becomes stronger as shown in the right panels of Fig.~\ref{fig:tr_bound_DFSZ}.
In the case of the SOA benchmark, the saxion decay into gravitino becomes very large for large saxion mass 
because of the large $\mu$ term.
Thus, the region of $m_{\ta}=m_s/2\gtrsim70$ TeV is excluded for all $T_R$.

\begin{figure}
\begin{center}
\includegraphics[width=7.5cm]{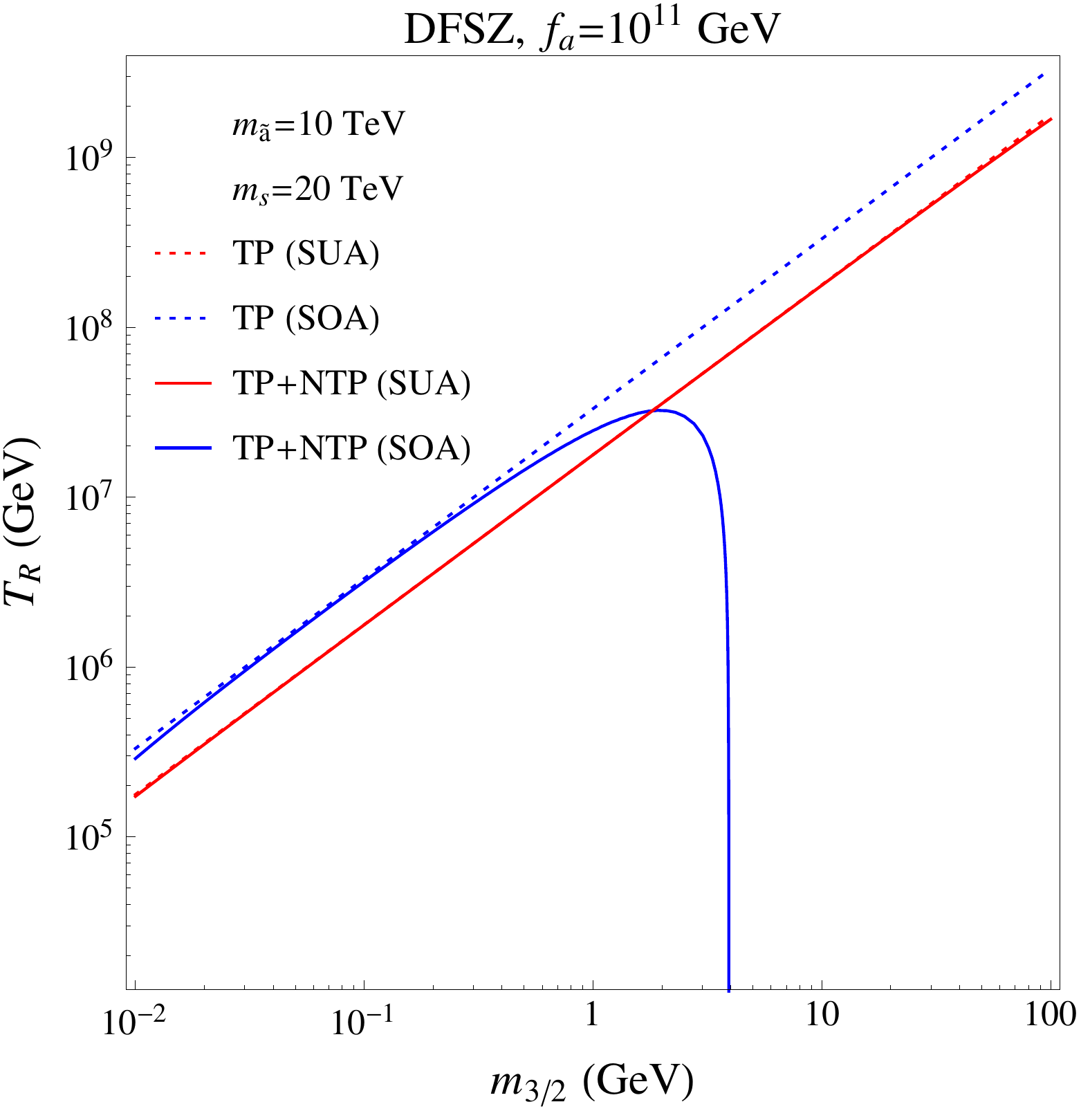}
\end{center}
\caption{The upper bound of $T_R$ calculated from 
thermal and non-thermal production of gravitinos as a function of $m_{3/2}$ for SUA (red) and SOA (blue) in the DFSZ model. 
The region above the curves is disallowed by overproduction of gravitinos.
\label{fig:tr_bound_DFSZ_gra}}
\end{figure}

In Fig.~\ref{fig:tr_bound_DFSZ_gra}, the upper bound of $T_R$ for varying $m_{3/2}$ is shown.
As discussed in the previous paragraphs, the gravitino production from the decays of axino and saxion is much smaller than the overclosure bound since the thermal productions of axino and saxion are much more suppressed than those in the KSVZ case.
Therefore, the $T_R$ bound is determined by the thermal production of gravitino.
In the case of SOA, meanwhile, the gravitino production from neutralino decay exceeds the overclosure bound for $m_{3/2}\gtrsim4$ GeV as in the KSVZ case.
Thus, there is no allowed region in this case.

\subsection{Hybrid KSVZ+DFSZ model} 

This hybrid model can be motivated by the simultaneous resolution to the $\mu$ problem and the domain wall problem 
achieving $N_{DW}=|6-N_\Phi|=1$.
The cosmological properties of the axion/saxion become somewhat different from those in the KSVZ and DFSZ model as they have both the QCD anomaly and  $\mu$-term interactions:
\begin{eqnarray}
{\cal L}&\supset&-\frac{\sqrt{2}  g_s^2}{32\pi^2 f_a/N_{\Phi} }\int d^2\theta AW^bW^b+\int d^2\theta~ \mu\left(1+2B_1\theta^2\right)H_uH_d e^{-2A/v_{PQ}}+{\rm h.c.}
\label{eq:gen_int}
\end{eqnarray}
valid below $M_\Phi \sim f_a=\sqrt{2}v_{PQ}$ and above $m_{\ta,s}$.
It is worth noting that the first term is generated only by PQ anomaly of heavy vector-like quarks, $\Phi+\Phi^c$, while 
the contribution from the ordinary quarks in the loop is still suppressed by $(\mu /E)^2$ as in the DFSZ case.

Thermal production of the axino and saxion for $T_R\gtrsim8\pi^2\mu$ is predominantly determined by the first term in Eq.~(\ref{eq:gen_int})
 and thus the axino/saxion thermal yield is the same as in Eq.~(\ref{eq:axn_TP}).
On the other hand,  the axino decay is determined by the second term as in the DFSZ case
if $m_{\ta}\lesssim8\pi^2\mu$. Therefore,
the gravitino production from the axino decay is typically given by
\begin{eqnarray}
\Omega_{\widetilde{G}}^{\ta}h^2&=&
2.8\times10^8\times \left(\frac{m_{3/2}}{\rm GeV}\right)
BR(\ta\to a+\widetilde{G})^{\rm DFSZ}~Y_{\ta}^{\rm KSVZ},\nonumber\\
&\simeq&2.3\times10^{-4}N_{\Phi}^2\left(\frac{\rm GeV}{m_{3/2}}\right)\left(\frac{100\mbox{ GeV}}{\mu}\right)^2
\left(\frac{m_{\ta}}{\rm TeV}\right)^4\left(\frac{T_R}{10^8\mbox{ GeV}}\right).
\end{eqnarray}
As in the previous cases, the saxion produced by coherent oscillation can contribute significantly 
to the gravitino density for large $f_a$.
The gravitino density from the saxion CO decay is the same as  in Eq.~(\ref{eq:saxCO_grav}) for the KSVZ case.

\begin{figure}
\begin{center}
\includegraphics[width=7.5cm]{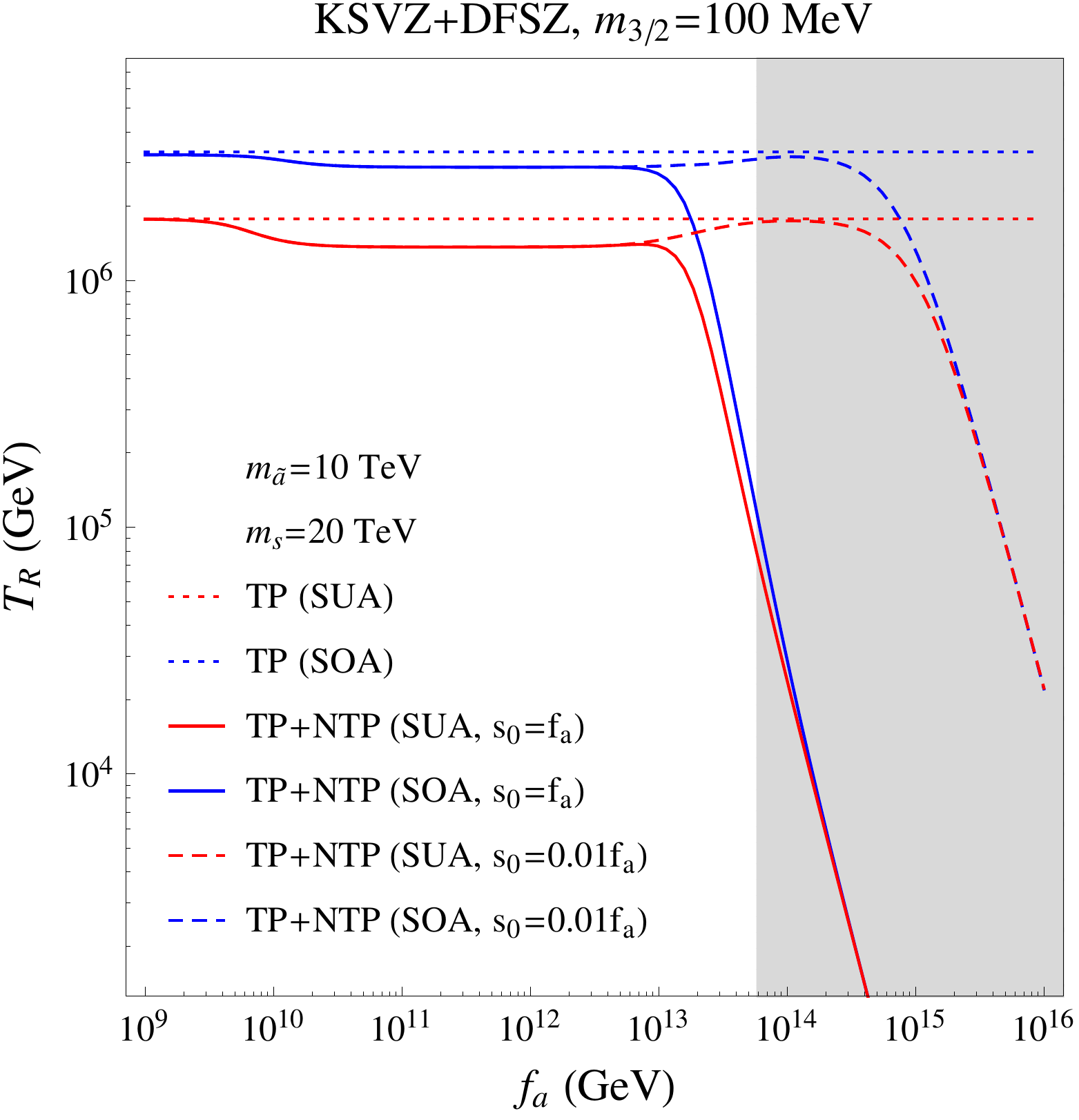}
\includegraphics[width=7.5cm]{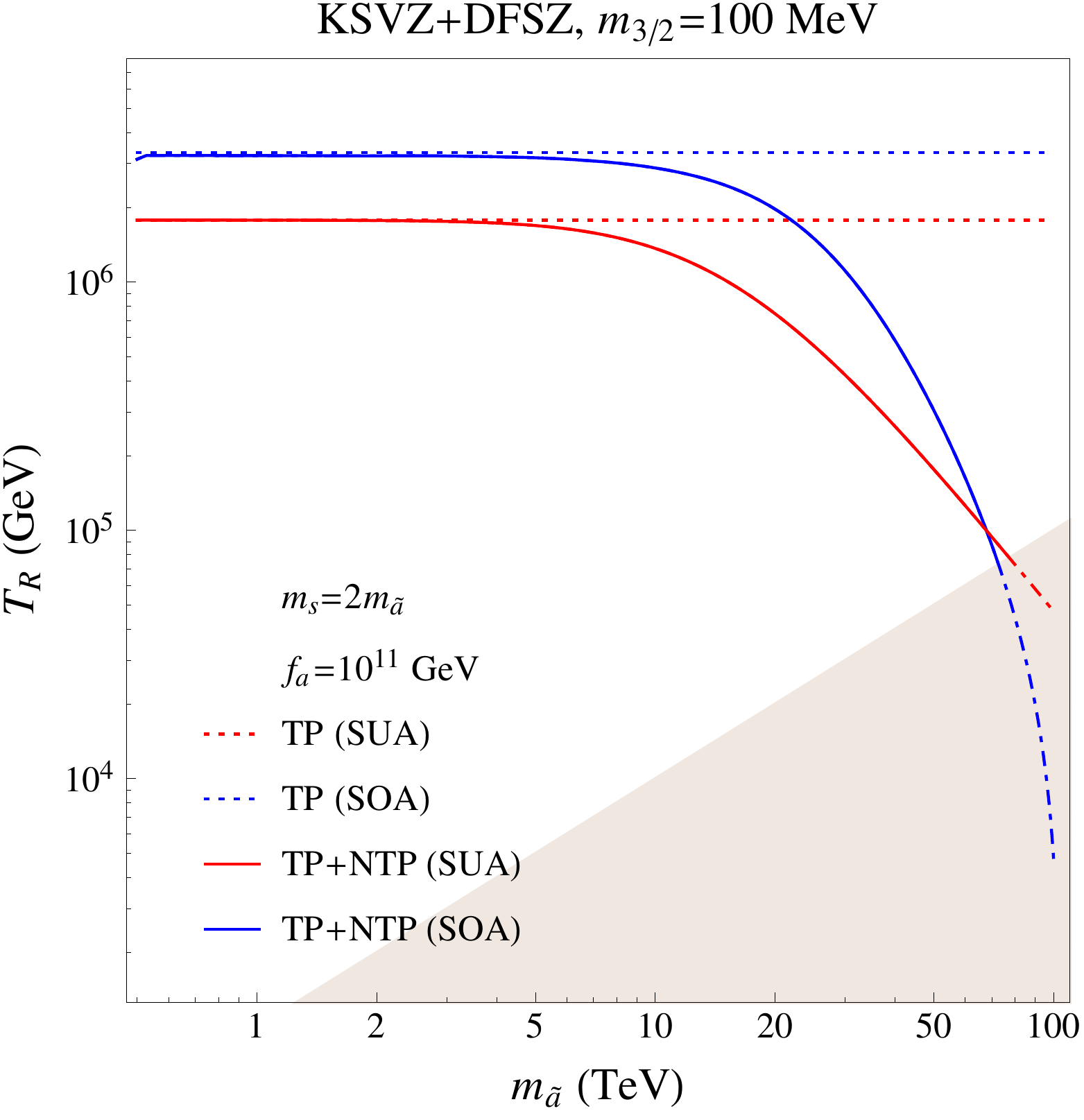}
\end{center}
\caption{The upper bound of $T_R$ calculated from 
thermal and non-thermal production of gravitinos as a function of $a)$ $f_a$ (left) 
and $b)$  $m_{\ta}$ (right) for SUA (red) and SOA (blue) in the KSVZ+DFSZ model. 
The region above the curves is disallowed by overproduction of gravitinos..
\label{fig:tr_bound_KSVZ+DFSZ}}
\end{figure}
\begin{figure}
\begin{center}
\includegraphics[width=7.5cm]{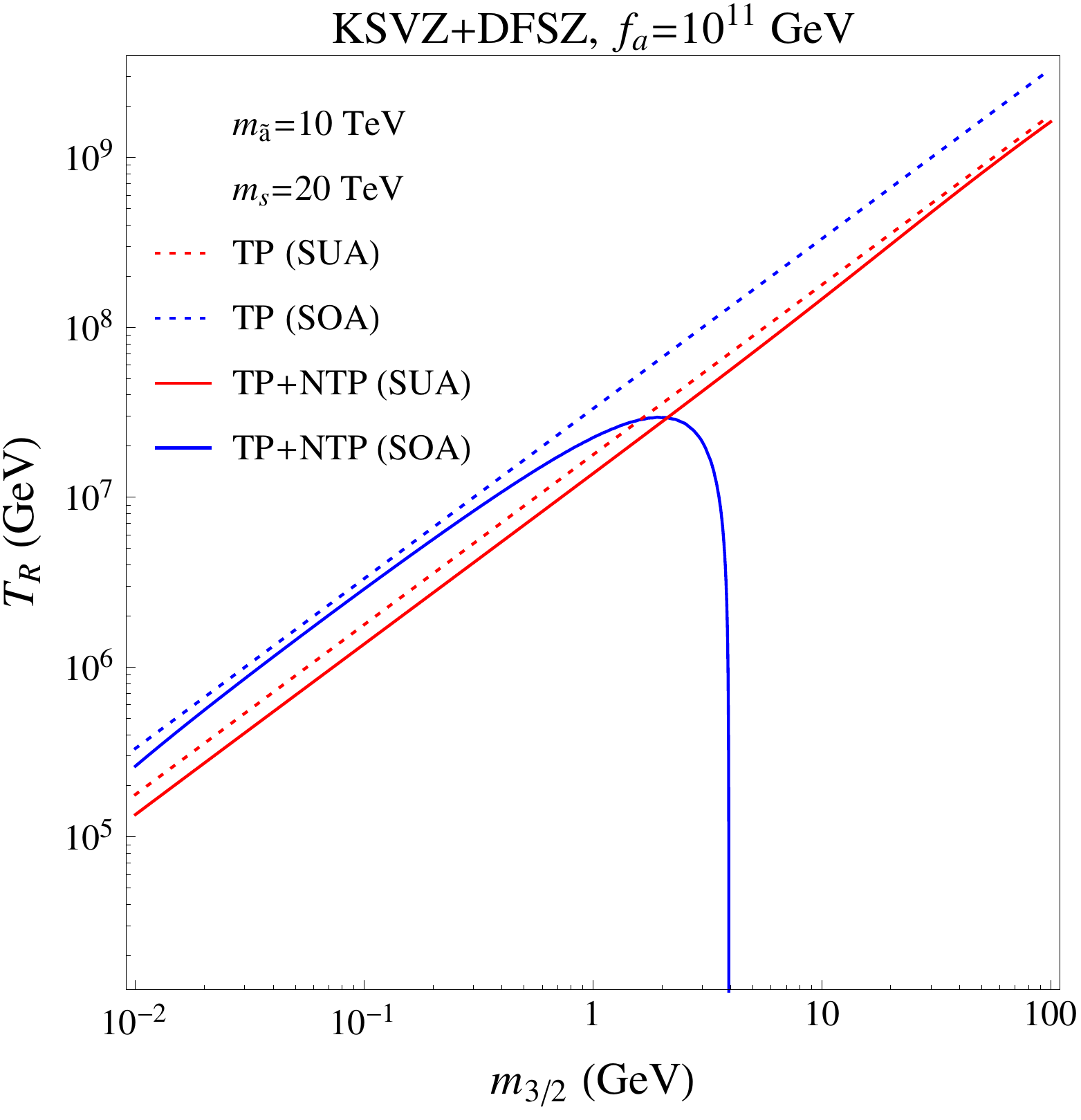}
\end{center}
\caption{The upper bound of $T_R$ calculated from 
thermal and non-thermal production of gravitinos as a function of $m_{3/2}$ for SUA (red) and SOA (blue) 
in the KSVZ+DFSZ model. 
The region above the curves is disallowed by overproduction of gravitinos.
\label{fig:tr_bound_KSVZ+DFSZ_gra}}
\end{figure}
%

In Fig.~\ref{fig:tr_bound_KSVZ+DFSZ}, we show the results of precise calculations
for the $T_R$ bounds depending on  $f_a$  and $m_{\ta}$ for the SUA and SOA benchmark points.
The main production of the axino and saxion is due to the anomaly interaction while the dominant decay is due to the Yukawa-type $\mu$-term interaction for $f_a\lesssim10^{13}$ GeV as we discussed.
The same amount of axinos and saxions are produced as in the KSVZ case, but they tend to decay more into MSSM particles so that $BR(\ta\to a+\widetilde{G})$ and $BR(s\to \ta+\widetilde{G})$ become smaller.
Therefore, the $T_R$ bounds are somewhere between those of the KSVZ and DFSZ cases.
For $f_a\gtrsim10^{13}$ GeV, the $T_R$ bound rapidly decreases because of the 
onset of saxion production via CO.
 If $s_0=0.01f_a$, saxion CO contribution becomes important for $f_a\gtrsim10^{15}$ GeV as in the KSVZ and DFSZ cases.
For production of gravitinos from neutralino production and decay, 
the production arguments are similar to those presented earlier for the KSVZ and DFSZ cases.


As in the cases of pure KSVZ or DFSZ, the upper bound of $T_R$ shows similar pattern which is shown in Fig.~\ref{fig:tr_bound_KSVZ+DFSZ_gra}.
There are sizable gravitino productions from both the thermal process and the axino/saxion decays, so the $T_R$ bound is slightly smaller than that from the thermal-only case.
Also, in the case of SOA benchmark, there is no allowed parameter space for $m_{3/2}\gtrsim4$ GeV because of the too much gravitino density from the neutralino decay.

\subsection{Long-lived neutralino and BBN}
In the gravitino LSP scenario, neutralino production and decay might result 
in post-BBN energy injection that disrupts the expected abundance of light elements. 
Depending on the life-time and decay modes, the neutralino abundance at the time of decay is constrained  as discussed in \cite{Kawasaki:2004qu,Jedamzik:2005sx}.  For $m_{\widetilde{Z}_1}={\cal O}(100 \mbox{ GeV})$, the bound on $\Omega_{\widetilde{Z}_1}h^2$ is given as
\bea
(B_h = 0.3)&&\quad\Omega_{\widetilde{Z}_1}h^2 \lesssim
\left\{\begin{array}{l l } 0.1 &{\rm for}\   \tau_{\widetilde{Z}_1}=1\sim100\sec\\
                                  4\times 10^{-4} &{\rm for}\  \tau_{\widetilde{Z}_1}=10^3\sec \\    
                                  (0.4\sim 1.0)\times 10^{-4} &{\rm for}\   \tau_{\widetilde{Z}_1} = 10^4 \sim 10^7\sec\\
                                  1.3\times 10^{-5} &{\rm for}\   \tau_{\widetilde{Z}_1} = 10^8 \sim 10^{12}\sec
                                  
                                  \end{array}\right., \label{eq:BBN1}\\
(B_h = 10^{-3}) &&\quad \Omega_{\widetilde{Z}_1}h^2 \lesssim
\left\{\begin{array}{l l } 40\sim 30 \hskip 1.8cm &{\rm for}\  \tau_{\widetilde{Z}_1}=1\sim 20\sec\\
                                   800 &{\rm for}\  \tau_{\widetilde{Z}_1} =40\sec\\
                                  0.1   &{\rm for}\  \tau_{\widetilde{Z}_1}=10^3\sec \\    
                                  0.01  &{\rm for}\ \tau_{\widetilde{Z}_1} = 10^4 \sim 10^6\sec \\
                                  10^{-4}  &{\rm for}\ \tau_{\widetilde{Z}_1} =10^7\sec \\
                                   10^{-5} &{\rm for}\ \tau_{\widetilde{Z}_1}=10^8\sim 10^{12}\sec
                                  \end{array}\right. \label{eq:BBN2}, 
\eea
where $B_h$ is the hadronic branching ratio which is crucial for 
$\tau_{\widetilde{Z}_1}  < 10^7\sec$. 
Here we took the conservative constraints on $^6$Li$/^7$Li as $^6$Li$/^7$Li $<$ 0.66. 
When the less conservative bound on $^6$Li$/^7$Li ($^6$Li$/^7$Li $<$ 0.1)  is used, 
the constrains become about eight times stronger in the range of $10^4\sim 10^6\sec$. 

Decay modes for the neutralino are given as   
(\ref{eq:neut_dec1}), (\ref{eq:neut_dec2}), and (\ref{eq:neut_dec3}).
The life-time and the relic abundance of the lightest neutralino for the SUA case (Higgsino-like) are
\begin{equation}\label{eq:Life_Neu}
\tau_{\widetilde{Z}_1}\simeq 1.7\times 10^2\sec\left(\frac{m_{3/2}}{100\, \rm MeV}\right)^2,\quad 
\Omega_{\widetilde{Z}_1}h^2 \simeq 0.013,
\end{equation}
and for the SOA case (Bino-like)
\begin{equation}\label{eq:Life_Neu}
\tau_{\widetilde{Z}_1}\simeq 1.2\times 10\sec\left(\frac{m_{3/2}}{100\, \rm MeV}\right)^2,\quad 
\Omega_{\widetilde{Z}_1}h^2 \simeq 6.8.
\end{equation}
The mode $\widetilde{Z}_1\to\widetilde{G}+Z$ is the main decay channel for both cases.
It is noted that the life-time in SUA case is about ten times longer than that in SOA case,
because the lightest neutralino of the SOA benchmark scenario has a sizable mixing component $v_4^{(1)}$ as $0.99$ 
compared to that of the Higgsino-like lightest neutralino ($v_4^{(1)}\sim 0.2$). 

When the decay modes $\widetilde{Z}_1\to \widetilde{G}+Z/h$ is sizable as for 
our benchmark points, the hadronic branching ratio is ${\cal O}(1)$. 
The value of $B_h$ can be suppressed if $m_{\widetilde{Z}_1}-m_{Z}  < m_{3/2}$ 
so that the neutralino decay to $Z/h$ is not kinematically allowed.  
However, in such a case where the life-time of $\tilde Z_1$ exceeds $10^8\sec$,
the constraints are mainly determined by electromagnetic decay and 
are still serious for $\Omega_{\widetilde{Z}_1}h^2 \lesssim 10^{-5}$.

The $\widetilde{Z}_1$ life-time can be shorter when $m_{3/2}$ is smaller.
For the SUA (SOA) benchmark point ($\Omega_{\widetilde{Z}_1}h^2= 0.013 (6.8)$),
$\tau_{\widetilde{Z}_1} $ has to be shorter than $200\sec$ ($0.12\sec$)
for $B_h = 0.3$ which implies that  $m_{3/2} \lesssim 100\,{\rm MeV}$ ($10\,{\rm MeV}$).

For $m_{3/2} > {\cal O}(100\, {\rm MeV})$, the decaying neutralino might be dangerous.
A way out is to consider Dirac neutrinos whose masses come from the Dirac Yukawa term:
\bea
W_{\nu} = y_\nu L N H_u, 
\eea
where $N$ is the right-handed (RH) neutrino superfield and  
$y_\nu$ is of ${\cal O}(10^{-13})$.  Here the conserved lepton number can be identified with the PQ symmetry so that
the smallness of $y_\nu$ might be explained by a nontrivial PQ charge of $N$ leading to  $y_\nu \sim (v_{PQ}/M_P)^n$.

 A special feature of the Dirac neutrino model is that the soft scalar mass of the 
RH sneutrino $m_{\widetilde N}$ is mostly dominated by gravity mediation 
due to a negligible contribution from gauge mediation. 
Thus $m_{\widetilde{N}}= O(m_{3/2}) < m_{\widetilde{Z}_1}$.
In this case,  $\widetilde{Z}_1$ decays mostly to $\tilde N + \nu$.
Then, we get \bea
\tau_{\widetilde{Z}_1}  
&\simeq& \left( \frac{(v_1^{(1)})^2 y_\nu^2 }{16\pi}\, m_{\widetilde{Z}_1} \right)^{-1}
= 3.9\times 10\sec \left(\frac{10^{-13}}{y_\nu}\right)^2
\eea
for the SUA case,
which is small enough to avoid the BBN constraint.
Non-thermally produced sneutrinos from neutralino decay will in turn decay to gravitinos via 
$\widetilde{N}\to \tilde G+\nu$ which does not affect the BBN.

On the other hand, in case of SOA, introducing the Dirac neutrino sector is not quite helpful for $m_{3/2} \gtrsim 10\,{\rm MeV}$ since the benchmark value $v_1^{(1)}$ is quite small. 
The dilution of the neutralino abundance via additional entropy production might be most promising in this region.

\section{Conclusion}

In this paper we have investigated cold dark matter production in supersymmetric axion
models characterized by the mass hierarchy $m_{3/2}\ll m_{\tz_1}\ll m_{\ta,s}$. In such models, 
the dark matter is expected to be composed of two particles: the axion $a$ and the gravitino $\tG$.
Whereas typically one might expect $m_{\ta}\sim m_s\sim m_{3/2}$ in gravity-mediation, we
derive a formal bound of $m_{\ta,s}<m_{3/2}(M_P/v_{PQ})$ which allows instead for
a heavy axino/saxion with $m_{\ta,s}\gg m_{3/2}$.

In the SUSY KSVZ model with heavy axino and gravitino LSP,  gravitinos are produced 
thermally with a relic abundance $\propto T_R$.
Gravitinos are also produced non-thermally
due to thermal production followed by decays of axinos in the early universe, and also by
thermal or CO production followed by decay of saxions into $\ta+\tG$. 
In this case, the thermal production of $\ta /s$ is also proportional to $T_R$.
In addition, gravitinos are produced due to neutralino production followed by 
(possibly late) decay to gravitinos. In this scenario, the neutralinos can be produced thermally,
or non-thermally themselves via axino or saxion production followed by decays.
The gravitino abundance is dominantly determined by the axino decay for the 
small $f_a$ region, while it is determined by the saxion decay for large $f_a$ if it is open.
We have seen that in the large $f_a$ and/or large $m_{\ta}$ region, 
the $T_R$ bound steeply decreases compared to that only from the thermal production.
In the KSVZ model, the suppressed decays of axinos, saxions and neutralinos must all 
be carefully evaluated in light of bounds from BBN on late decaying semi-stable relics.

In the SUSY DFSZ model, the direct coupling of the
axion superfield to the Higgs superfields leads to
axino/saxion thermal production rates which are independent of
$T_R$ so that the upper bound on $T_R$ is mainly determined by
the thermal production of gravitinos . Furthermore, in the SUSY DFSZ case,
axino and saxion decays tend to be more rapid than in the KSVZ case
(for a given value of axino/saxion masses).
This latter condition implies that
the saxions and axinos tend to decay before the onset of BBN,
and often even before neutralino freeze-out. For these reasons,
SUSY DFSZ axion models with a gravitino as LSP are much less constrained
than corresponding models with a KSVZ axion.


As consequences of this scenario with mixed axion/gravitino dark matter, we expect 
ultimate detection of relic axions, but {\it no} detection of WIMP dark matter.
However, we would still expect detection of supersymmetric particles at
colliding beam experiments, given sufficiently energetic beams and increased integrated luminosity.

\section*{Acknowledgments}
We thank the US Department of Energy for support for this research 
project. C.S.S. is supported in part by DOE grants doe-sc0010008, 
DOE-ARRA- SC0003883, and DOE-DE-SC0007897.

\end{document}